\documentclass[10pt]{iopart}
\usepackage{iopams}

\expandafter\let\csname equation*\endcsname\relax
\expandafter\let\csname endequation*\endcsname\relax
\usepackage{amsmath}

\usepackage{float}
\usepackage{graphicx}
\usepackage{dcolumn}
\usepackage{bm}
\usepackage{color}
\usepackage{soul}
\usepackage{multirow}
\usepackage[colorlinks,plainpages=false,linkcolor=blue,urlcolor=blue,citecolor=blue,pdfpagemode=UseNone,pdfstartview=FitBH]{hyperref}
\usepackage[square,numbers,sort&compress]{natbib}
\newcommand{\YbAl}{YbAlO$_3$}
\newcommand{\YFe}{YFeO$_3$}
\newcommand{\DySc}{DyScO$_3$}
\newcommand{\TbFe}{TbFeO$_3$}
\newcommand{\DyFe}{DyFeO$_3$}
\newcommand{\HoFe}{HoFeO$_3$}

\newcommand{\TmFe}{TmFeO$_3$}
\newcommand{\YbFe}{YbFeO$_3$}

\makeatletter
\g@addto@macro\@floatboxreset\centering
\makeatother

\begin{document}

\title[Spin dynamics in perovskite oxides]{Low-energy spin dynamics in rare-earth perovskite oxides}

\author{A~Podlesnyak$^1$, S~E~Nikitin$^2$ and G Ehlers$^3$}

\address{$^1$ Neutron Scattering Division, Oak Ridge National Laboratory, Oak Ridge, Tennessee 37831, USA}
\address{$^2$ Paul Scherrer Institut, CH-5232 Villigen, Switzerland}
\address{$^3$ Neutron Technologies Division, Oak Ridge National Laboratory, Oak Ridge, Tennessee 37831, USA}

\eads{\mailto{podlesnyakaa@ornl.gov}, \mailto{stanislav.nikitin@psi.ch}, \mailto{ehlersg@ornl.gov}}
\vspace{10pt}
\begin{indented}
\item[]June 2021
\end{indented}

\begin{abstract}
We review recent studies of spin dynamics in rare-earth orthorhombic perovskite oxides of the type $RM$O$_3$, where $R$ is a rare-earth ion and $M$ is a transition-metal ion, using single-crystal inelastic neutron scattering (INS).
After a short introduction to the magnetic INS technique in general, the results of INS experiments on both transition-metal and rare-earth subsystems for four selected compounds (\YbFe, \TmFe, \YFe, \YbAl) are presented.
We show that the spectrum of magnetic excitations consists of two types of collective modes that are well separated in energy: gapped magnons with a typical bandwidth of $<$70~meV, associated with the antiferromagnetically (AFM) ordered transition-metal subsystem, and AFM fluctuations of $<$5~meV within the rare-earth subsystem, with no hybridization of those modes.
We discuss the high-energy conventional magnon excitations of the 3$d$ subsystem only briefly, and focus in more detail on the spectacular dynamics of the rare-earth sublattice in these materials.
We observe that the nature of the ground state and the low-energy excitation strongly depends on the identity of the rare-earth ion.
In the case of non-Kramers ions, the low-symmetry crystal field completely eliminates the degeneracy of the multiplet state, creating a rich magnetic field-temperature phase diagram.
In the case of Kramers ions, the resulting ground state is at least a doublet, which can be viewed as an effective quantum spin-1/2.
Equally important is the fact that in Yb-based materials the nearest-neighbor exchange interaction dominates in one direction, despite the three-dimensional nature of the orthoperovskite crystal structure.
The observation of a fractional spinon continuum and quantum criticality in \YbAl\ demonstrates that Kramers rare-earth based magnets can provide realizations of various aspects of quantum low-dimensional physics.
\end{abstract}

%
\vspace{2pc}

\vspace{1pc}
\noindent{(Some figures may appear in colour only in the online journal)}

%
\maketitle
%
\ioptwocol

\section{Introduction}

Almost two centuries ago, in 1839, German mineralogist Gustav Rose discovered in the Ural Mountains of Russia a calcium titanium oxide mineral CaTiO$_3$, which later was named after Russian scientist Lev Perovski~\cite{Katz}.
This finding opened an era of scientific research on perovskite, a broad family of crystals that share the same structural arrangement as the mineral CaTiO$_3$~\cite{Goldschmidt,Pena}.
A perovskite structure can contain almost every element from the Periodic Table~\cite{Filip}.
This extraordinary diversity of perovskite results in a fascinating array of physical properties, and led to countless studies over the last century.

Nowadays, the rare-earth perovskite oxides $RM$O$_3$ with orthorhombic structure (where $R$ is a rare-earth ion, Bi or Y, and $M$ is a transition-metal ion) are an important family of materials whose properties remain a focus of considerable attention due to many intriguing physical phenomena they demonstrate.
Remarkable examples include, but are not limited to, multiferroic \cite{Cheong,Khomskii} and ferroelectric~\cite{Cohen} properties, temperature and field induced spin reorientation transitions \cite{Belov1974,BelovBook}, laser-pulse induced ultrafast spin-reorientation~\cite{Kimel, Jong, Jiang2013}, anisotropic magnetic entropy evolution~\cite{Ke2016}, magneto-optical effects \cite{Kimel}, exotic quantum states and quantum spin dynamics at cryogenic  temperatures~\cite{Nikitin2018,Wu2019PRB,Wu2019Nat,Agrapidis,Rong2020}.
The rich magnetic phase diagram of these materials arises from peculiarities of the 3$d$ and 4$f$ magnetic sublattices and the interplay between them.
In $RM$O$_3$ with a magnetic 3$d$ ion, a strong Heisenberg superexchange interaction $M$-$O$-$M$ induces a robust antiferromagnetic (AFM) order at several hundreds of degrees Kelvin ~\cite{Yamaguchi1973,Shapiro1974,Yamaguchi1974,Hahn}.
In some of $RM$O$_3$, in addition to the AFM order, a Dzyaloshinskii-Moriya (DM) interaction causes a spin canting giving rise to a weak ferromagnetic (FM) moment.
With decreasing temperature or in applied magnetic field, in a number of orthoperovskites the 3$d$ magnetic sublattice undergoes a spin reorientation transition~\cite{White1969,Yamaguchi1974}.
In the case of non-magnetic $M=$~Al, Sc, or Co (in its low-spin state~\cite{Jirak2014}), the magnetic properties of $RM$O$_3$ are primarily controlled by an electronic structure of rare-earth ion $R^{3+}$ and rare-earth inter-site interactions.
In turn, the crystalline electric field (CEF) splits the lowest-lying $4f$ multiplet and determines the magnetic single-ion anisotropy as well as the magnitude of the magnetic moment.
The rare-earth sublattice only orders at a few Kelvin, if at all, indicating much weaker exchange coupling between the 4$f$ moments~\cite{White1969}.
The interaction between the two spin subsystems (3$d$ and 4$f$) also plays an important role and often determines the magnetic ground state.
For instance, the interplay between the Fe and Dy sublattices gives rise to gigantic magneto-electric phenomena in \DyFe~\cite{Tokunaga}.

\begin{figure}[b!]
	\includegraphics[width=.8\linewidth]{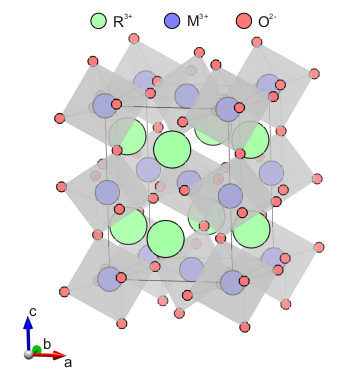}
	\caption{~Schematic crystal structure of orthorhombic perovskites, $RM$O$_3$. Green, blue and red balls indicate rare-earth, transition metal and oxygen ions respectively.}
	\label{Fig:structure_introduction}
\end{figure}

The details of the ground state of the 4$f$ magnetic sublattice remain poorly understood and the $R$-$R$ coupling at low temperatures is often not even considered ~\cite{Yamaguchi1974,Belov1976,Bazaliy2005}.
It is generally accepted that quantum effects are strongest with the smallest possible spin. Large anisotropic rare-earth $4f$ moments with strong spin-orbit coupling (SOC) are considered to be classical, since Heisenberg exchange interaction~\cite{Heisenberg} cannot reverse their directions~\cite{Wu2016Orbital}.
However, this is correct in the context of single-ion physics, and the crystalline environments may change the electronic properties of the rare-earth ions considerably.
The energy levels of the $R^{3+}$ with an odd number of electrons in the unfilled 4$f$ shell (Kramers ions) are split by the CEF into doubly degenerate states.
The quantum states of the ground Kramers doublet, separated by a large energy gap from the first excited state, are the superpositions of its ``up'' and ``down'' components, and therefore the doublet can be viewed as an effective quantum spin-1/2~\cite{Onoda,Wu2016Orbital}.
In the case of non-Kramers ions, the CEF splitting may result in singlets, doublets, pseudo-doublets, or states with higher degeneracy, depending on the symmetry.
Accordingly, one can expect rich magnetic field-temperature phase diagrams for compounds with non-Kramers rare-earth ions.

This review focuses on recent studies of spin dynamics in rare-earth perovskite oxides.
They revealed an intriguing coexistence of the classical high-energy spin waves of the transition metal magnetic sublattice and unconventional low-energy spin excitations of the rare-earth sublattice, which spontaneously transform from classical magnon to quantum quasiparticles with temperature and magnetic field~\cite{Nikitin2018,Wu2019PRB,Wu2019Nat,Agrapidis,Rong2020}.
We show that in spite of its three-dimensional perovskite structure, $RM$O$_3$ can provide
realizations of various aspects of a quantum spin $S = 1/2$  quasi-one-dimensional chain, enabling the observation of a quantum fractional spinon continuum~\cite{mourigal2013}.
Because of the wide variety of perovskite materials, it is hardly possible to attempt even a short overview of the entire $RM$O$_3$ family.
Therefore we intentionally skip many other systems with exciting properties, like multiferroic manganites, relaxor ferroelectrics, etc.
We encourage the interested reader to take a look at the many excellent reviews on the topic, e.g.~\cite{Pena,Dong,Cowley,Khomskii,Xiang,Fu,Liu}.
We focus on the properties of four $RM$O$_3$ orthoperovskites, aiming to cover various possible combinations of the magnetic - nonmagnetic sublattices in the system: i) \YFe\ with a nonmagnetic $R$ site; ii) \YbAl\ with a nonmagnetic $M$ site; iii) \YbFe\ with both magnetic sublattices and a Kramers rare-earth ion; iv) \TmFe\ with both magnetic sublattices and a non-Kramers rare-earth ion.

The  presentation  is  organized  as  follows: we start with a description of the basic
principle of inelastic neutron scattering (INS) and neutron scattering instruments (section~\ref{ins}).
In section~\ref{scales} we discuss a hierarchy of the magnetic energy scales and corresponding exchange interactions in rare-earth perovskites.
section~\ref{3d} contains a short review of spin dynamics in transition-metal subsystem.
section~\ref{rare-earth}  describes a single-ion and collective magnetic behavior of the rare-earth ion sublattice.
Here, we emphasise the unusual quantum properties of the rare-earth subsystem arising from strong quantum fluctuations at low temperature.
Finally, section~\ref{outlook} presents our conclusion remarks.

\section{Inelastic neutron scattering}
\label{ins}

\begin{figure}[t!]
	\includegraphics[width=1\linewidth]{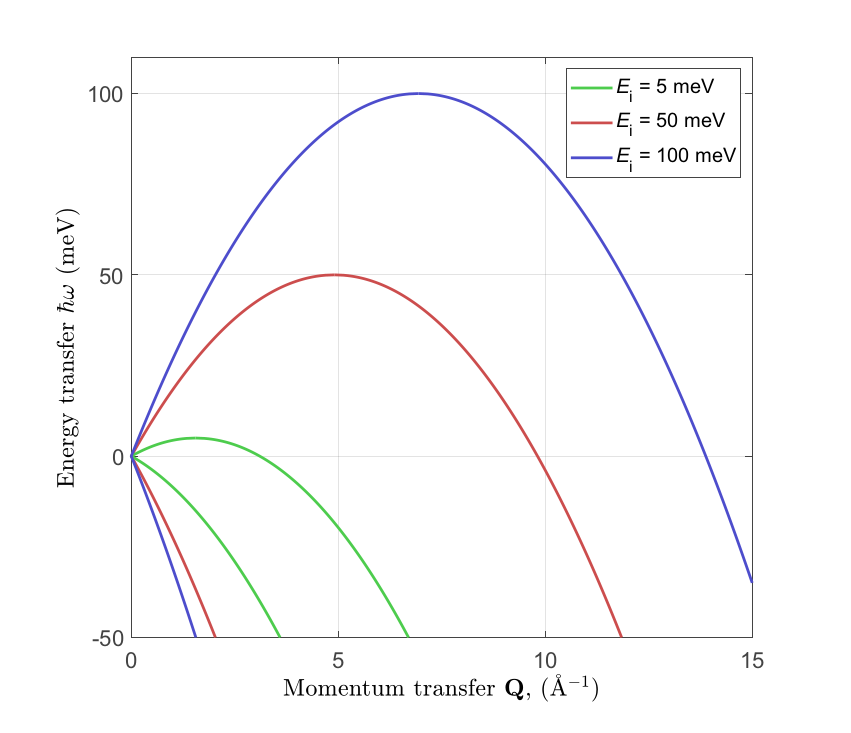}
	\caption{~Dynamical range for \emph{direct geometry} inelastic neutron scattering spectrometer calculated for three incident neutron energies. The angular detector coverage ranging from 0 to 180$^{\circ}$.}
	\label{Exp:DynamicalRange}
\end{figure}

Neutron scattering is one of the most suitable tools to study magnetic and lattice structures and excitations in crystalline materials~\cite{Gurevich1968,Izumov,ShiraneShapiro,Squires78, Lovesey86}.
The strengths of this method are directly linked to the physical properties of neutrons.
The neutron is an elementary particle that has mass, and a spin moment $S = 1/2$, but no electrical charge\footnote{With the precision of the up-to-date experiments. However, the experimental search for the neutron electric dipole moment remains a challenging problem in nuclear physics.}.
Therefore, the neutron interacts with nuclei and with magnetic moments in condensed matter. Unlike scattering probes such as electrons or photons, the neutron does not directly interact with localized or itinerant charges.
Moreover, due to the absence of an electric charge, neutrons can deeply penetrate into the solids and thereby, probe the bulk properties, whereas most other $\mathbf{Q}$-resolved spectroscopic techniques, such as angle-resolved photoemission spectroscopy (ARPES) or resonant inelastic x-ray scattering (RIXS), are surface probes.

In crystalline materials, the magnetic and phonon excitations have a typical energy scale of $\hbar\omega\approx1$~meV -- 0.1~eV.
We are interested in studying their dispersion within several Brillouin zones, which normally implies a transferred momentum of $\mathbf{Q} \approx$ [0.1--10]~\AA$^{-1}$.
The wave vector of the neutron $\mathbf{k}$ is related to its energy by the standard equation for the kinetic energy of a massive particle, $E=(\hbar^2\mathrm{k}^2)/(2m_{\mathrm{n}})$.
The actual value of the neutron mass $m_{\mathrm{n}}$ enables one to cover the desired ranges of transferred energies and momenta simultaneously, see figure~\ref{Exp:DynamicalRange}.

\begin{figure}[t!]
	\includegraphics[width=0.8\linewidth]{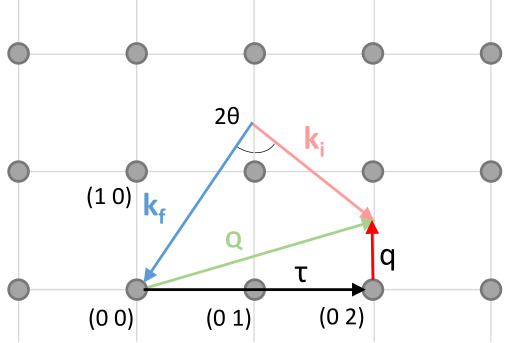}
	\caption{~Visualization of equation \eqref{exp:Conserv1} in reciprocal space for inelastic neutron scattering. }
	\label{ExpScatteringDiagram}
\end{figure}

In the scattering process, the neutron changes its energy and momentum, and we can write the conservation laws:
\begin{align}
&\hbar\mathbf{Q} = \hbar(\mathbf{k}_{\mathrm{i}} - \mathbf{k}_{\mathrm{f}}) 	\label{exp:Conserv1}\\
&\hbar\omega = E_{\mathrm{i}} - E_{\mathrm{f}} = \frac{\hbar^2}{2m_{\mathrm{n}}}
(\mathrm{k}^2_{\mathrm{i}} - \mathrm{k}^2_{\mathrm{f}}), \label{exp:Conserv2}
\end{align}
where $\mathbf{k_{\mathrm{i}}}, \mathbf{k}_{\mathrm{f}}$ and $E_{\mathrm{i}}$, $E_{\mathrm{f}}$ are momentum and energies of incoming and outgoing neutron, respectively; $\mathbf{Q}$ and $\hbar\omega$ are momentum and energy transferred between neutron and scattering entity.
For $|\mathbf{k_{\mathrm{i}}}| = |\mathbf{k}_{\mathrm{f}}|$ we have from \eqref{exp:Conserv2} $\hbar\omega = 0$, which is commonly referred to as ``elastic scattering'' (the neutron energy and wavelength are not changed).
Inelastic scattering is visualized in figure~\ref{ExpScatteringDiagram} where the scattering vector can be decomposed according to $\mathbf{Q} = \mathbf{\tau} + \mathbf{q}$ with $\mathbf{q}$ being the wavevector of an elementary excitation within the first Brillouin zone, and $\mathbf{\tau}$ is a reciprocal lattice vector.
Neutron scattering is thus able to measure very directly a dispersion relation $\hbar\omega(\mathbf{q})$ at any predetermined point in reciprocal space.

The measured quantity in the neutron scattering experiment is the count rate $C$ in the detector. This value is proportional to the total neutron flux incident on the sample, the efficiency of the detector, etc; however, the most important quantity, which enters $C$ and characterizes the interaction of neutrons with the sample, is the double differential scattering cross-section $d^2\sigma/(d\Omega d\omega)$.
This cross-section defines the number of neutrons that are scattered into an element of solid angle $d\Omega$ with energy between $E_{\mathrm{f}}$ and $E_{\mathrm{f}}+ dE_{\mathrm{f}}$.

Due to the interaction with the neutron, the scattering system changes its state from $\lambda_{\mathrm{i}}$ to $\lambda_{\mathrm{f}}$.
Using Fermi's Golden Rule and the Born approximation we can write down the master equation~\cite{ShiraneShapiro, Lovesey86}:
\begin{align}
	\frac{d^2\sigma}{d\Omega d\omega}  &=
       \frac{\mathbf{k}_{\mathrm{f}} } {\mathbf{k}_{\mathrm{i}}}
	 \left(\frac{m_{\mathrm{n}}}{2\pi\hbar^2}\right)^2 		
	 \sum_{\lambda_{\mathrm{i}}} p_{\lambda_{\mathrm{i}}} \sum_{\lambda_{\mathrm{f}}}
	 |\langle \mathbf{k}_{\mathrm{f}}, \lambda_{\mathrm{f}} | \mathrm{\hat{U}} |
	  \mathbf{k}_{\mathrm{i}}, \lambda_{\mathrm{i}}\rangle|^2 \nonumber \\
	 &\times \delta(\hbar\omega + E_{\mathrm{i}} - E_{\mathrm{f}}).
\label{Exp:CrossSec1}		
\end{align}
Here, $\lambda_{\mathrm{i}}$ denotes the initial state of the scatterer, with energy $E_{\mathrm{i}}$ and thermal population factor $p_{\lambda_{\mathrm{i}}}$, and its final state is ${\lambda_{\mathrm{f}}}$.
$\mathrm{\hat{U}}$ is the interaction operator of the neutron with the sample, which
 can be written as
\begin{equation}
	\mathrm{\hat{U}} =  \sum_l V_l (\mathbf{r} - \mathbf{r}_l(t)),
\label{exp:interactionoperator}
\end{equation}
where $\mathbf{r}_l$ is the position of the scattering objects in the sample.

\subsection{Nuclear scattering.}
Interaction of the neutrons with nuclei takes place on very short distances, much shorter than the de Broglie wavelength of thermal neutrons.
Therefore, we can consider the scattering potential as a delta-function at the nuclei positions $\delta(\mathbf{R} - \mathbf{r}_j)$.
Therefore, after proper mathematical transformations~\cite{ShiraneShapiro} we can rewrite \eqref{Exp:CrossSec1} in the form
\begin{equation}
	\frac{d^2\sigma}{d\Omega d\omega}  = \frac{\mathbf{k}_{\mathrm{f}} }{\mathbf{k}_{\mathrm{i}}}
	N b^2 S(\mathbf{Q},\omega),
\label{exp:NuclearCrossSec}
\end{equation}
where $N$ is the number of nuclei in the scattering system.
In this equation we introduced two important quantities.
The first one is the scattering length $b$ that quantifies the scattering potential of the particular nucleus at $\mathbf{r}_l(t)$ with the neutron.
Note, that the neutron scattering cross-section can be subdivided into coherent and incoherent parts, and the total cross-section is their sum
	$\sigma_\mathrm{total} = \sigma_\mathrm{coh}  + \sigma_\mathrm{incoh}$.
Pure coherent scattering would only exist in a material with a single type of equivalent scattering centers.

In most materials, coherent and incoherent nuclear scattering originate from the fact that the actual scattering length at a site $\mathbf{r}_l(t)$ is random for two reasons.
Different isotopes of an element are randomly distributed over the crystallographic sites, and have different individual scattering lengths.
Moreover, a nuclear spin also contributes to the scattering length, and is usually disordered.
The coherent and incoherent parts of the cross-section can be expressed using the scattering length $b$ as $\sigma_\mathrm{coh}  = 4\pi \langle b \rangle^2$ and $\sigma_\mathrm{incoh} = 4\pi (\langle b^2 \rangle - \langle b \rangle^2$).
The values of the scattering length for different isotopes are tabulated~\cite{sears1992neutron}.

\begin{figure*}[tb]
\includegraphics[width=0.8\linewidth]{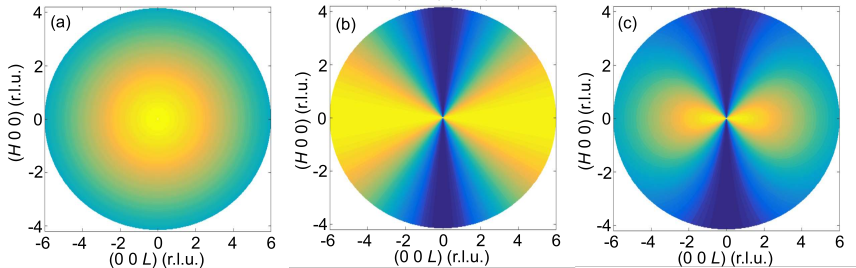}
	\caption{~Calculated scattered intensity for the longitudinal fluctuations of Yb moments in the $(H0L)$ plane of \YbFe. Panel (a) shows the effect of the magnetic form-factor; (b)~effect of the polarization factor; (c)~their combination.}
	\label{Exp:ExpPolarization}
\end{figure*}

The second important term in \eqref{exp:NuclearCrossSec} is the scattering function $S(\mathbf{Q},\omega)$ defined as the Fourier transform of the pair correlation function \cite{VanHove}:
\begin{equation}
	S(\mathbf{Q},\omega) = \frac{1}{2 \pi \hbar} \int
	G(\mathbf{r},t) e^{i (\mathbf{Q} \cdot \mathbf{r} - \omega t)}
	{\;}d\mathbf{r} dt,
\end{equation}
\begin{equation}
	G(\mathbf{r},t) = \frac{1}{(2\pi)^3} \frac{1}{N} \int \sum_{l, l'}
	\langle e^{-i\mathbf{Q} \cdot \mathbf{r}_{l'}(0)} e^{i\mathbf{Q} \cdot \mathbf{r}_l(t)} \rangle {\;}d\mathbf{Q}.
	\label{Exp:corfunc}
\end{equation}
This function contains information regarding both static and dynamical properties of the sample, and usually its determination is the aim of a neutron scattering experiment.
It has two important properties,
\begin{align}
	&S(\mathbf{Q},\omega) = \mathrm{exp}(\frac{\hbar\omega}{k_{\mathrm{B}}T}) S(-\mathbf{Q},-\omega) {\;},		\label{Exp:Balance}  \\
	&S(\mathbf{Q},\omega)   = \frac{\chi''(\mathbf{Q},\omega)}{1 - \mathrm{exp}(- \frac{\hbar\omega}{k_{\mathrm{B}}T}) }{\;}. \label{Exp:DisFluct}
\end{align}
Here, \eqref{Exp:Balance} represents the principle of detailed balance.
It shows that the probability for the creation of an excitation is proportional to $\langle n+1 \rangle$ and that the destruction of an excitation is proportional to $\langle n \rangle$, where
\begin{equation}
	\langle n \rangle = \frac{1}{\mathrm{exp}(\frac{\hbar\omega}{k_{\mathrm{B}}T}) + 1}{\;}.
\end{equation}
Equation~\eqref{Exp:DisFluct} is a consequence of the fluctuation-dissipation theorem and shows the connection between the scattering function and the imaginary (dissipative) part of the dynamical susceptibility $\chi''(\mathbf{Q},\omega)$ of the scattering system.
It allows a direct comparison of $\chi''(\mathbf{Q},\omega)$ as obtained by neutron scattering with bulk measurements of magnetic susceptibility.

\subsection{Magnetic scattering.}
\label{magscatt}

Since neutrons have a spin moment, they also interact with magnetic moments of unpaired electrons.
In this case the scattering potential $V$ can be considered as a simple interaction between two magnetic dipoles.
In the simplest case, an interaction between the electron moving with momentum $\mathbf{p}$ and spin $\mathbf{s}$ and the neutron can be expressed as:
\begin{equation}
	V = -2 \gamma \mu_{\mathrm{B}} m_{\mathrm{n}} \sigma (\Delta \times \frac{\mathbf{s}\times\mathbf{r}}{|r|^3} + \frac{\mathbf{p}\times\mathbf{r}}{\hbar |r|^3}),
\end{equation}
where $\sigma$ represents the spin moment of the neutron.
We substitute the defined interaction potential into \eqref{Exp:CrossSec1} and without going into details (see for example~\cite{Lovesey86,Squires78,Zaliznyak2005,JensenMackintosh_book1991}) obtain the double-differential cross-section for magnetic scattering:
\begin{eqnarray}
\frac{d^2\sigma}{d\Omega d\omega} \propto &|F(\mathbf{Q})|^2 \sum_{\alpha,\beta} \frac{g_\alpha}{2} \frac{g_\beta}{2} \left(\delta_{\alpha \beta} -\frac{Q_\alpha Q_\beta}{Q^2} \right) \sum_{{l,l'}} e^{i\mathbf{Q \cdot r_{\textit{ll'}}}} \nonumber \\
\times \sum_{\lambda_{\mathrm{i}}, \lambda_{\mathrm{f}}} & p_{\lambda_{\mathrm{i}}} \langle\lambda_{\mathrm{i}}|\hat{S}_{l}^{\alpha}|\lambda_{\mathrm{f}}\rangle \langle\lambda_{\mathrm{f}}|\hat{S}_{l'}^{\beta}|\lambda_{\mathrm{i}}\rangle \delta(\hbar\omega + E_{\lambda_{\mathrm{i}}} - E_{\lambda_{\mathrm{f}}}). \;\;\;\;\;\;\;\;
\label{exp_magnetic_cross_section}
\end{eqnarray}
Here, $\alpha, \beta = x, y, z$ and $g_\alpha$, $g_\beta$ are the corresponding $g$-factors; $\hat{S}_{l}^{\alpha}$ and $\hat{S}_{l'}^{\beta}$ are the angular momentum operators at sites $l$ and $l'$, respectively.

Equation~\eqref{exp_magnetic_cross_section} has two important terms.
The first one is the magnetic form factor $F(\mathbf{Q})$, which is defined as the Fourier transform of the magnetization density.
$F(\mathbf{Q})$ can be expressed in terms of the expectation values of spherical Bessel functions  $j_l(|\mathbf Q|)$~\cite{ShiraneShapiro}.
In the most common dipole approximation (which is valid for small $|\mathbf{Q}|$) it consists of two terms:
\begin{equation}
F(|\mathbf{Q}|)=\langle j_0(|\mathbf{Q}|) \rangle + \frac{2-g}{g}\langle j_2(|\mathbf{Q}|) \rangle.
\end{equation}
The coefficients $j_0$ and $j_2$ are tabulated~\cite{Brown}.
Since the form factors for the magnetic and structural excitations are different, analyzing the $|\mathbf{Q}|$ dependencies of an excitation we may understand whether it has magnetic or lattice origin.

The second term $(\delta_{\alpha, \beta} - Q_{\alpha} Q_{\beta} / Q^2)$ is the so-called polarization factor of the magnetic neutron scattering.
It takes into account the fact that only spin components perpendicular to $\mathbf{Q}$ contribute to the magnetic scattering cross-section.
This is a very useful property because it provides information on the orientation of the magnetic moments, especially in the case of collinear magnets.

To illustrate both the magnetic form factor and the polarization factor, let us consider an ensemble of magnetic moments (Yb$^{3+}$ in our case) oriented along the $a$-axis of an orthorhombic sample.
The calculated scattering pattern in the $(H0L)$ plane taking into account only the form factor is shown in figure~\ref{Exp:ExpPolarization}(a).
One can see that the intensity has a rotational symmetry and decreases with $\mathbf{Q}$ unlike to the phonon scattering (which increases with $\mathbf{Q}$).
Figure~\ref{Exp:ExpPolarization} (b) shows the polarization factor, which varies with the angle between scattering vector and magnetic moment.
Figure~\ref{Exp:ExpPolarization} (c) shows the intensity that would be observed, which is the product of both, polarization and the form factor, and one can see that the polarization factor suppresses the intensity along the moment direction.
Such an analysis allows one to determine the directions of the magnetic moments in a crystal
when the moments have only preferred orientation without long-range ordering, as in the case of \YbFe\ or \TmFe\ (see section~\ref{Chp:TmFeO3}).
Note, that the calculated pattern is valid for elastic scattering, and the longitudinal part of the inelastic spectrum.
In the case of transverse excitations, such as spin-waves, the polarization factor has a different form~\cite{Lovesey86}.

\subsection{Time-of-flight spectrometer.}

\begin{figure}[tb]
	\includegraphics[width=1\linewidth]{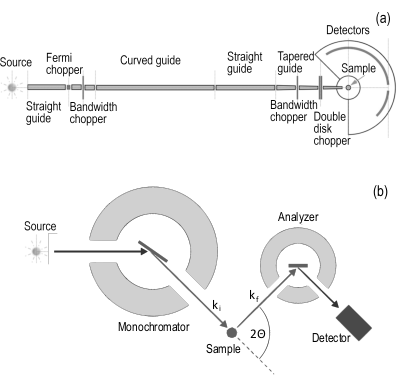}
	\caption{~(a)~General layout of the time-of-flight spectrometer CNCS at the Spallation Neutron Source, Oak Ridge National Laboratory. (b) Sketch of the triple-axis spectrometer.
}
	\label{Exp:Spectrometers}
\end{figure}

In order to obtain $S(\mathbf{Q},\omega)$, we need instrumentation to access the variables $\mathbf{Q}$ and $\omega$ throughout reciprocal space.
Two commonly used instrument types for this purpose are time-of-flight (TOF) neutron spectrometers and triple-axis (TAS) spectrometers\footnote{Bertram Brockhouse was awarded a share of the 1994 Nobel Prize in Physics for the development of neutron spectroscopy and the TAS instrument in particular.}~\cite{Brockhouse}, see figure~\ref{Exp:Spectrometers}.

The basic idea of a TOF spectrometer is rather simple: we illuminate a sample with a monochromatic beam of neutrons\footnote{For simplicity, here we describe only a direct geometry spectrometer. For more details about inverse geometry spectrometers see, e.g.~\cite{carpenter_loong_2015}}, which is subdivided in time into short pulses. When interacting with the sample, neutrons change their momenta and energies (and as a consequence, their velocity). After the scattering process, the neutrons fly to the detectors, and we register their arrival time at the detector. Knowing the flight path, time of detection and the initial velocity of the neutron, we can calculate how much energy it lost or gained during the interaction with the sample. Moreover, analyzing the angle where the neutron was detected, we can also work out the transferred momentum.
The general layout of the TOF Cold Neutron Chopper Spectrometer (CNCS) \cite{CNCS1,CNCS2} at the Spallation Neutron Source (SNS), Oak Ridge National Laboratory (ORNL), where most of the INS data discussing in this review were obtained, is shown in figure~\ref{Exp:Spectrometers}(a).
One of the main strengths of TOF spectrometers is that we can simultaneously measure a full spectrum of the transferred energies $\hbar\omega < E_{\mathrm{i}}$. Moreover, we can cover a large solid angle with the detectors, and therefore, measure a rather large portion of the four-dimensional $\mathbf{Q}$-$\omega$ space in one scan.
Usually nowadays the measurements are done using the rotating crystal method, where the crystal is rotated by 360$^{\circ}$  with a step of $0.5 - 2^{\circ}$. After data collection, one can use a standard software, such as \textsc{Mantid}~\cite{Mantid}, \textsc{Dave}~\cite{AzuahKneller09} or \textsc{Horace}~\cite{Horace} to convert the data into energy-momentum coordinates and symmetrize them according to the crystal symmetry.

\subsection{Triple-axis spectrometer.}
The TAS is the most versatile instrument for INS because it can be configured in many ways and is more flexible with resolution than a TOF instrument~\cite{ShiraneShapiro}.
The exceptional convenience of a TAS spectrometer is that the measurement can be done at pre-determined points in reciprocal space (so called constant-$\mathbf{Q}$ scan) or for a fixed energy transfer $\hbar \omega$ (constant-$E$ scan).
By performing multiple scans it is possible to map out the complete dispersion.
Figure~\ref{Exp:Spectrometers}(b) shows schematics of a TAS spectrometer.
Incident neutrons with a well defined initial wave vector $\mathbf{k}_{\mathrm{i}}$ are selected from the white spectrum of the neutron source by the monochromator crystal (first axis) and scattered from the sample (second axis).
The crystal analyzer (third axis) reflects the scattered neutrons with the wavevector $\mathbf{k}_{\mathrm{f}}$ onto the neutron detector defining the energy transfer $\hbar \omega$.

\section{Magnetic energy scales}
\label{scales}

\begin{figure*}[t!]
  \includegraphics[width=1\linewidth]{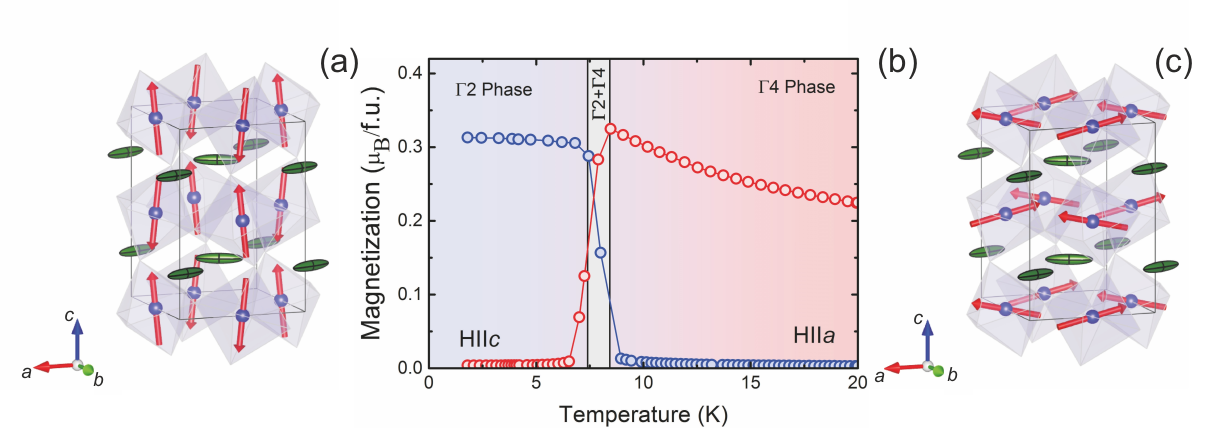}
  \caption{~Sketches of magnetic structures of \YbFe\ below (a) and above (c) the SR transition. Blue spheres show Fe ions, green ellipsoids represent anisotropic magnetic moments of Yb. In the magnetic phase $\Gamma2$ (a), below $T_{\mathrm{SR}}$, Fe moments align along the $c$ axis, and spin canting results in a net moment along the $a$ axis. Above $T_{\mathrm{SR}}$ ($\Gamma4$ phase), Fe moments rotate to the $a$ axis, and spin canting gives a net moment along the $c$ axis. (b)~Temperature dependencies of the magnetization of \YbFe\ measured at $B = 0.01$~T along the $c$ (red) and $a$ (blue) axes. Reproduced with permission from~\cite{Nikitin2018}.
}
\label{YBFO_Structure}
\end{figure*}

The rare-earth orthorferrites $R$FeO$_3$ are an important family of materials whose magnetic properties remain a focus of considerable research.
Here, on the example of orthoferrites, we shortly discuss the  hierarchy of the magnetic energy scales and corresponding exchange interactions.

The transition-metal sublattice consists of Fe$^{3+}$ ions, which have a $3d^5$ electronic configuration with spin moment of $S =5/2$ and zero orbital momentum.
The moments are coupled by strong $\sim180^{\circ}$ superexchange interaction via Fe-O-Fe path, and thus one would expect robust 3D magnetic order with negligible magnetocrystalline anisotropy. Indeed, the iron subsystem in $R$FeO$_3$ materials orders into a $\mathrm{k} = 0$ canted AFM structure $\mathrm{\Gamma}4$ at high temperature with $T_{\mathrm{N}} \approx 600$~K, and the spin canting gives a weak net FM moment along the $c$ axis, which can be detected in magnetization measurements (see figure~\ref{YBFO_Structure})~\cite{White1969, bozorth1958magnetization, Bazaliy2005}.
Careful neutron diffraction measurements have found additional canting along the $b$-axis, which is symmetric relative to the $ac$-plane and does not produce a net moment~\cite{plakhtij1981experimental}.
The canting was found to be of the order of $\approx1^{\circ}$ and is the result of antisymmetric Fe-Fe DM exchange interactions, which are allowed in the system due to low symmetry.

With temperature decreasing, a spontaneous spin reorientation (SR) transition from $\mathrm{\Gamma}4$ to a $\mathrm{\Gamma}2$ or $\Gamma1$ magnetic configuration occurs in many orthoferrites with magnetic $R$-ions~\cite{White1969, bozorth1958magnetization} in a wide temperature range from $T_{\mathrm{SR}}\approx450$~K for SmFeO$_3$ down to $T_{\mathrm{SR}}\approx7.6$~K for \YbFe, and the net magnetic moment rotates from the $a$ to the $c$ axis [see figure~\ref{YBFO_Structure}].
It was shown that the SR transition is caused by the competition between anisotropies of the rare-earth and transition-metal sublattices.
A consistent microscopic treatment of $3d$-$4f$ interaction in these systems is extremely complex, because of low symmetry and number of terms in the magnetic Hamiltonian~\cite{Yamaguchi1973,Yamaguchi1974}.
However, the rare-earth and Fe sublattices can be described phenomenologically by considering the free energy of the system~\cite{BelovBook,Bazaliy2004spin,Bazaliy2005}.
The spin reorientation was considered in great detail previously, and in this review we will not focus on this topic, but rather recommend references~\cite{Yamaguchi1973,BelovBook}, which cover the main approach to this long-standing issue.

As a general rule, in insulators the $4f$ electrons have more compact orbitals and thus exhibit much weaker superexchange interaction than $3d$ ions.
Accordingly, the ordering in the rare-earth sublattice, associated with $R-R$ interaction, in orthorhombic perovskites takes place at a lower temperature, $T \approx 1-10$~K, if at all~\cite{White1969}.

However, the key difference between the magnetic behavior of the  transition-metal and rare-earth subsystems comes from the fact that in the first case the SOC is weak and the orbital momentum is quenched.
Therefore, the magnetocrystalline anisotropy is small compared to the exchange interaction energy scale (with an exception of Co$^{2+}$, which shows similar behavior with $4f$ ions~\cite{Raveau}).
In contrast, the strong spin-orbit coupling of the rare-earth ions entangles the $S$ and $L$ quantum numbers into a total angular momentum $J = |L \pm S|$ (the sign depends on the sign of the SOC constant, $\lambda_{\mathrm{SOC}}$) and the energy scale of the splitting between multiplets with different $J$ is on the order of $0.1-1$~eV.
Thus, the low-temperature physics is entirely determined by the lowest multiplet.

The CEF causes a further splitting within each $J$ multiplet.
We note that in the case of so-called Kramers ions with odd number of $f$ electrons, the electric field cannot break the time reversal symmetry, $\hat{T}$, and the CEF leads to $(2J+1)/2$ Kramers doublets.
Each doublet is constructed of the wavefunctions $|\psi_1\rangle, |\psi_2\rangle$, which obey $|\psi_1\rangle = \hat{T} |\psi_2\rangle$.
For ions with even number of $f$-electron (integer momentum), there is no such protection and in general case the multiplet $J$ can be split into $2J + 1$ singlet states.
The typical splitting caused by the CEF is on the scale of $0.1-100$~meV.
Thus, we have a strong hierarchy of energy scales, $E_{\mathrm{SOC}} > E_{\mathrm{CEF}} > E_{\mathrm{Exchange}}$ and the low-energy magnetic properties are predominately dictated by several low-energy states provided by CEF effects.

\begin{figure}[tb!]
  \includegraphics[width=1\linewidth]{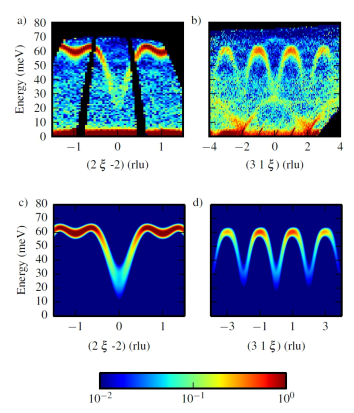}
  \caption{~(a) Measured spin wave dispersion in \YFe\ along a) (2, $\xi$, -2) and (b) (3, 1, $\xi$) and calculated spin-wave dispersion along (c) (2, $\xi$, -2) and d) (3, 1, $\xi$). The excitation with double periodicity at lower energy is a phonon mode. Black pixels show regions where no data were collected. The measurements were done using the Fine Resolution Chopper Spectrometer (SEQUOIA)~\cite{SEQ} at the SNS at ORNL. Reproduced with permission from~\cite{Hahn}.
}
  \label{YFO_Magnon}
\end{figure}

Considering the relevant energy scales in rare-earth orthoferrites, we might expect several different types of magnetic excitations at different energy scales:
(i) robust AFM magnons within the transition-metal subsystem and magnon bandwidth of $T^{\mathrm{Fe}}_{\mathrm{N}}/k_{\mathrm{B}} \sim 50-80$~meV;
(ii)  collective excitations due to $R$-$R$ correlations at $\sim 1$~meV;
(iii) complex magnons due to coupling between $R$ and transition-metal subsystems at intermediate temperatures;
(iv) intra-atomic localized CEF or inter-multiplet excitations in $R$ ions at $1-100$~meV.

\section{Short review of spin dynamics in the transition-metal subsystem}
\label{3d}

Comprehensive investigations of the spin dynamics in the  transition-metal subsystem were mainly focused on the Fe$^{3+}$ sublattice in orthoferrites~\cite{Shapiro1974, Gukasov1997, Hahn, Park2017low, Nikitin2018, Skorobogatov2020low,Ovsyanikov2020neutron}.
Here we give a quick overview of those studies.

\begin{figure}[t!]
  \includegraphics[width=1\linewidth]{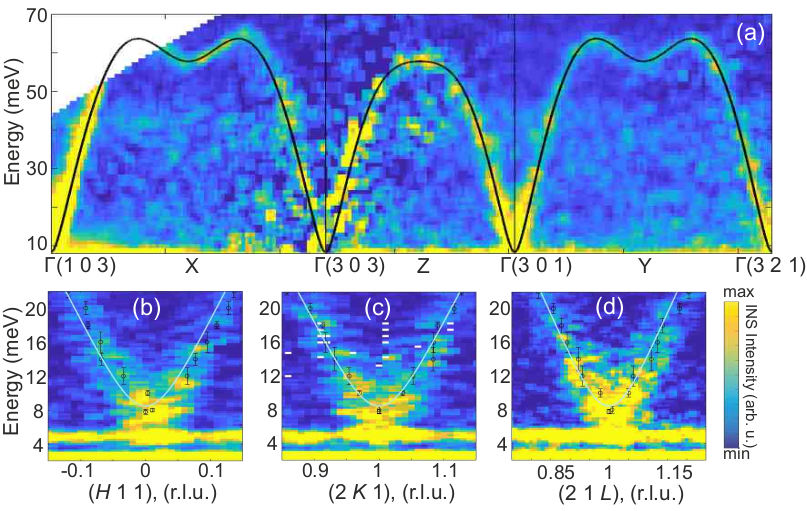}
  \caption{~INS spectra of \TmFe\ measured at $T = 7$~K.
  (a)~Energy-momentum cut through the high-symmetry directions measured with $E_{\mathrm{i}}~=~100$~meV using the wide angular-range chopper spectrometer (ARCS) \cite{ARCS} at the SNS at ORNL.
  (b-d)~Energy-momentum cuts along $H, K$ and $L$ directions close to the $\Gamma$ point of the magnetic Brillouin zone show the magnon gap. The spectra were taken with $E_{\mathrm{i}}~=~25$~meV.
  The solid lines shown in all panels represent the results of the LSWT calculations. Reproduced with permission from~\cite{Skorobogatov2020low}.
  }
\label{TFO_MagnonFig1}
\end{figure}

The INS experiments have shown, that the high energy excitations of  all studied orthoferrites look similar with respect the energy scale and bandwidth of the magnons.
As an example, figures~\ref{YFO_Magnon} and \ref{TFO_MagnonFig1} show the experimental INS spectra taken from \YFe\ and \TmFe, respectively.
Observed magnon branches with an energy scale of $E\approx60$~meV stem from the magnetic Bragg peaks with an even sum of $H + K + L$, and are clearly associated with collective excitations of the Fe$^{3+}$ magnetic moments.

\begin{figure}[t!]
  \includegraphics[width=0.9\linewidth]{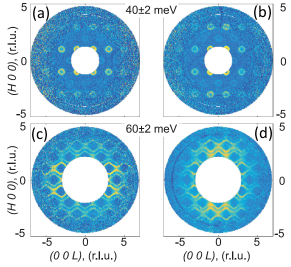}
  \caption{~Constant energy slices of the INS intensity in \YbFe\ within the $(H0L)$ scattering plane taken at $T=2$~K (a,c) and $T=15$~K (b,d). The scattering intensities were integrated within energy windows that are indicated between the corresponding panels. The measurements were done using the SEQUOIA spectrometer at the SNS at ORNL. Reproduced with permission from~\cite{Nikitin2018}.
}
  \label{const-en}
\end{figure}

Figure~\ref{const-en} presents constant-energy slices in the ($H0L$) plane taken from \YbFe\ around energies $E = 40$, 60~meV at $T = 2$~K (left) and $T = 15$~K (right), below and above $T_{\mathrm{SR}} = 7.6$~K.
The slices show the clean spin-wave excitations caused by the Fe-Fe interaction for both temperatures, and one can see the redistribution of the INS intensity, which is concentrated either along the $H$ or $L$ direction, at $T = 2$ and 15~K, respectively, as expected from
the known SR transition of the Fe moments.

The Fe spin fluctuations can be reasonably well described using a simple linear spin-wave theory (LSWT)~\cite{Toth}.
To describe the magnetic structure and spin dynamics of the Fe subsystem the general Hamiltonian can be written in the following form~\cite{Hahn,Skorobogatov2020low}:
\begin{eqnarray}
 \mathcal{H}_{\rm Fe} =  \sum_{i,j} \ {\mathbf{S}^{\rm Fe}_{i} \cdot \hat{J}_{ij} \cdot \mathbf{S}^{\rm Fe}_{j}}  - \sum_i \mathbf{S}^{\rm Fe}_i \cdot K_i \cdot \mathbf{S}^{\rm Fe}_i,
 \label{Eq:FeHam1}
\end{eqnarray}
where the first term describes exchange interaction between different pair of Fe ions, and the second term shows effect of quadratic single-ion anisotropy.
We further note that the exchange matrix is non-diagonal and includes antisymmetric DM terms,
\begin{eqnarray}
\hspace{1cm} \hat{J} =
	\biggl(
		\begin{array}{ccc}
					J_{xx}&  D_{xy} & -D_{xz} \\
				   -D_{yx}&  J_{yy} &  D_{yx} \\
					D_{zx}& -D_{zy} &  J_{zz}
	\end{array}
	\biggr).	
 \label{trash}
\end{eqnarray}

\begin{figure}[t!]
  \includegraphics[width=0.9\linewidth]{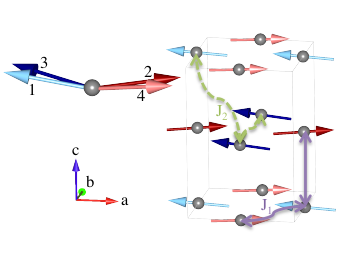}
  \caption{~Magnetic unit cell of \YFe, showing only the positions of the Fe$^{3+}$ atoms. The four sublattices show weak ferromagnetism and antiferromagnetism along the $c$ and $b$ directions, respectively. Exchange interactions between nearest ($J_1$) and next-nearest ($J_2$) neighbors are shown by the solid purple and dashed green arrows, respectively. Reproduced with permission from~\cite{Hahn}.
}
\label{YFO_Structure}
\end{figure}

\begin{figure*}[tb!]
  \includegraphics[width=0.9\linewidth]{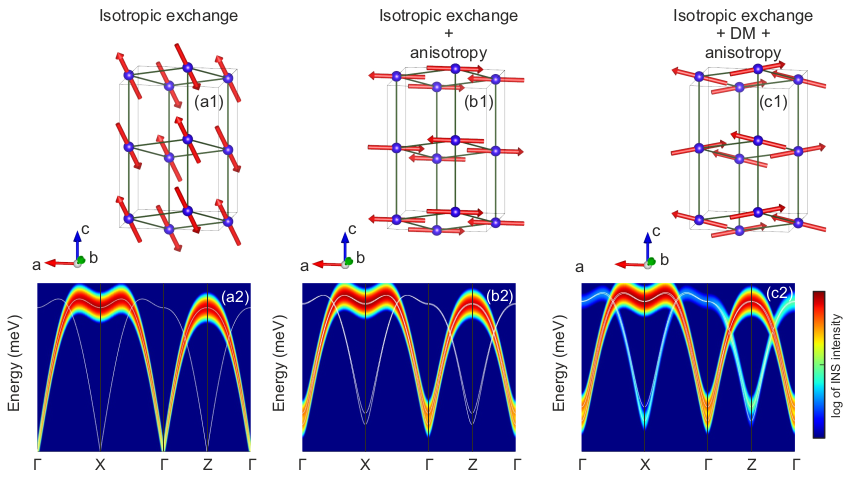}
  \caption{~Effect of different terms in Hamiltonian~\eqref{Eq:FeHam1} on the magnetic structure and excitation spectra. Magnetic structures of $R$FeO$_3$ in presence of isotropic exchange interaction,~(a1); isotropic exchange interaction and anisotropy, $K_yS_y^2 + K_xS_x^2$ ($|K_x|>|K_y|, K_x < 0)$,~(b1); isotropic exchange interaction, DM interaction and anisotropy,~(c1). (a2-c2) Magnon excitation spectra calculated in the presence of correspondent terms along two representative directions of the Brillouin zone. Note that anisotropy and DM terms are overemphasized in order to present their effect more clearly. The intensity in panels (a2-c2) is shown on logarithmic scale.
}
  \label{spectra_sketch}
\end{figure*}

$R$FeO$_3$ has four Fe ions in a unit cell, and thus one would expect four magnon modes.
The isotropic Heisenberg interaction is a leading term in \eqref{Eq:FeHam1}, and it generates two two-fold degenerate magnon modes, but only one of each carries spectral intensity.
Moreover, in order to obtain a proper description of spin waves one has to take into account exchange interactions within the second coordination sphere.
Due to the structural distortion, the exchange interactions within each coordination sphere are slightly ($\sim 10$~\%) anisotropic with $J_c > J_{ab}$~\cite{Park2017low, Skorobogatov2020low, Ovsyanikov2020neutron}.

The dominating uniaxial easy-axis anisotropy term in \eqref{Eq:FeHam1}, $K_xS_x^2$ (or $K_zS_z^2$), opens a magnon gap, while the second term, $K_yS_y^2$, lifts the degeneracy of two-fold degenerate modes.
To stabilize the correct four-sublattice magnetic ground state it is essential to include two DM terms, one of each, $D_{xz}$ causes the FM net moment and the second, $D_{yz}$ and $D_{xz}$ induces additional canting, which does not produce a net moment, although it  generates the proper four-sublattice structure as shown in figure~\ref{YFO_Structure}.
The DM terms influence the low-energy splitting of the modes as well. In addition, they cause a redistribution of the spectral intensity towards two modes, which would carry zero spectral weight otherwise.
However, due to the weakness of the DM terms these modes carry rather small spectral intensities and thus are hard to observe in an INS experiment.
The effects of the different terms on the magnon spectrum and magnetic ground state is represented schematically in figure~\ref{spectra_sketch}.

Now let us briefly discuss the influence of the $R$-Fe interaction on the high-energy magnons.
In the materials with a nonmagnetic ion at the $R$-site (Y, La, Lu, etc.), the spin gap is rather small ($\sim1-2$~meV), and has only a marginal temperature dependence between 1.5 and 300~K~\cite{Park2017low}.
This spin gap is caused by the actual single-ion anisotropy of the Fe$^{3+}$.
On the other hand, in the materials with a magnetic rare-earth ion, the value of the spin gap exhibits a strong temperature dependence and vanishes (or softens considerably) at the SR transition temperature~\cite{Shapiro1974,balbashov1989anomalies, dan1994energy, buchel1999relative}. In the vicinity of $T_{\mathrm{SR}}$ the system becomes unstable against external perturbations, and a laser-induced ultrafast spin reorientation was observed in several orthoferrites~\cite{Kimel, de2011laser}.
The $R$-Fe interaction renormalizes \textit{effective} anisotropy constants considerably and increases the gap, up to 8~meV in \TmFe\ at low temperature, $T \ll\ T_{\mathrm{SR}}$~\cite{Skorobogatov2020low}.
At high temperatures, the dominating term in the $\Gamma4$ phase is the easy-axis contribution of $K_c$, while in the SR region the $K_a$ becomes dominant, and thus causes a transition to a $\Gamma2$ phase.
Finally we note that, at the moment, there is no clear evidence for collective dispersive $R$-Fe excitations from INS experiments.

\section{Magnetism of the rare-earth subsystem}
\label{rare-earth}

\subsection{CEF Hamiltonian, point-charge model, symmetry of wavefunctions}
\label{cef-pcm}
Before we begin with the description of the collective magnetic behavior, it is essential to discuss the single-ion electronic state of a rare-earth ion in $RM$O$_3$. Due to the octahedral distortions, the symmetry of the $4c$ position, occupied by the $R$ ions, is significantly lowered from $O_h$ in the perfect cubic perovskite structure to $C_s$.
$R$ ions are surrounded by eight distorted nearest-neighbor $M$-O octahedra. The $R$ position has only one mirror plane, which is perpendicular to the $c$-axis, and therefore the magnetic symmetry implies that the moment can be either pointing along the $c$-axis or lie in the $ab$-plane. It is worth noting that in this structure, the $R$ ions occupy two symmetry-equivalent positions, which experience a CEF with identical strength and symmetry but rotated by $\pm\phi^{\circ}$ to the $a$-axis in the $ab$-plane, see a sketch in figure~\ref{Fig_anisotropy} (d).

The effective CEF Hamiltonian can be written as~\cite{Stevens1952}:
\begin{equation}
 \mathcal{H} = \sum_{l,m} B_{l}^{m}O_{l}^{m} {\;},
 \label{Eq:CEF1}
\end{equation}
where $B_l^m$ and $O_l^m$ are Stevens coefficients and operators.

Because the $R$ ions occupy the low-symmetry position $C_s$, the CEF Hamiltonian \eqref{Eq:CEF1} includes 15 independent $B_l^m$ parameters.
Therefore, the analysis and precise determination of the individual parameters is a complicated task, which requires a large number of observables (such as intensities and energies of CEF excitations, detailed measurements of the magnetic susceptibility, etc.).
To the best of our knowledge, a complete determination for an orthorhombic perovskite was done only for NdFeO$_3$~\cite{Przenioslo}.

Another method to evaluate $B_l^m$ parameters is to calculate them ab-initio and the simplest approach is a point-charge (PC) model~\cite{Stevens1952,Hutchings1964,Rotter2004,Rotter2011}.
The model considers only Coulomb forces and does not account for a possible screening or orbital hybridization. However, in the case of the ionic crystals with strongly localized $4f$ electrons it can provide a reasonable description of the CEF ground state.
Such calculations were successfully performed for cases with Kramers ions, namely \DySc~\cite{Wu2017DyScO3} and \YbAl~\cite{Wu2019PRB}.
The calculated CEF doublet states are best diagonalized when the local Ising axes are chosen along the Ising moment direction, which in these materials means along a direction at $\phi$ degrees titled from the easy axis.
Note, that although the PC CEF calculation is only an approximation, and the energy scheme cannot quantitatively reproduce the real ground state wave function, it does qualitatively confirm an important detail: in both cases there is a well separated ground state doublet, which is Ising-like with a majority contribution from the largest $M_z$ wave function,
$| \pm 7/2 \rangle$ (Yb) or $| \pm 15/2 \rangle$ (Dy).
This was further confirmed by the INS, specific heat and magnetization measurements.
Moreover, the PC CEF calculations were performed for \TmFe\ with non-Kramers rare-earth ion~\cite{Skorobogatov2020low}.
They correctly captured a $c$-axis Ising-like magnetic anisotropy of the ground state singlet, intensity and polarization of CEF excitations, although failed to reproduce quantitatively the transition energies.

\subsection{Spin chain physics in \YbAl}
\label{YbAlO3}

Perhaps the most unexpected result of our recent INS work on rare-earth based perovskites was an observation of the quasi-1D character of the 4$f$ subsystem.
The possible explanation of the origin of the quasi-1D character in spite of the 3D crystal structure can be as following. The combination of strong spin-orbit coupling and crystalline electrical field effects creates an energetically isolated Kramers doublet ground state.
The ground state doublet has a strong uniaxial anisotropy, which constrains the magnetic moments in the $ab$ plane with an angle $\alpha$ tilted out of the $a$ axis.
The intra-plane dipolar interaction, been the main coupling in the plane, depends strongly on the relative tilting angle of the Ising moments \cite{Kappatsch}.
It turns out that in a number of rare-earth perovskites, including \YbAl\ and \YbFe, the angle of the Ising moments is a ``magic'' number, at which the dipolar interaction vanishes \cite{Wu2019PRB}.
Thus, one-dimensional interaction along the $c$ axis becomes dominant and suggests \YbAl\ and \YbFe\ as new members of one-dimensional quantum magnets.
We also cannot exclude an unusual spatial anisotropy of the superexchange interaction, which is cancelled out for in-plane directions.
Detailed ab-initio calculations are required to resolve this question.
We describe the low-temperature magnetism in \YbAl\ as quantum critical Tomonaga-Luttinger liquid that features spinon confinement-deconfinement transitions in different regions of the magnetic field-temperature phase diagram~\cite{Wu2019Nat,Nikitin21Nat}.

\YbAl\ has been reported as an insulating antiferromagnet, which orders below  $T_{\mathrm{N}} \approx$ 0.8~K into a coplanar, but non-collinear antiferromagnetic (AFM) structure $AxGy$, with the Yb moments lying in the $ab$-plane forming an angle $\phi~=~28^{\circ}$ with the $a$-axis ($m_{\mathrm{Yb}} = 3.7 \pm 0.3~\mu_{\mathrm{B}}$)~\cite{Radhakrishna1981}.
The authors of \cite{Radhakrishna1981} proposed that the Yb moments form 1D Ising-like chains, and this model could partially explain their results. However, a full understanding of the magnetic behavior of the material was not achieved.
In particular, much essential information, such as the type of the magnetic anisotropy, and the hierarchy of magnetic interactions, could not be determined from their experiments because of the poly-crystalline nature of the sample.
The successful growth of large single crystal of \YbAl\ by the Czochralski technique~\cite{Buryy2010, Noginov2001} enabled qualitatively new investigations of the magnetic properties of this material.
In this section we review the results of recent experimental and theoretical work that was devoted to the low-temperature magnetism in this material~\cite{Wu2019PRB,Wu2019Nat,Agrapidis,Rong2020,Fan2020}.

\subsubsection{Introduction and general characterization.}

\begin{figure}[tb!]
  \includegraphics[width=1\linewidth]{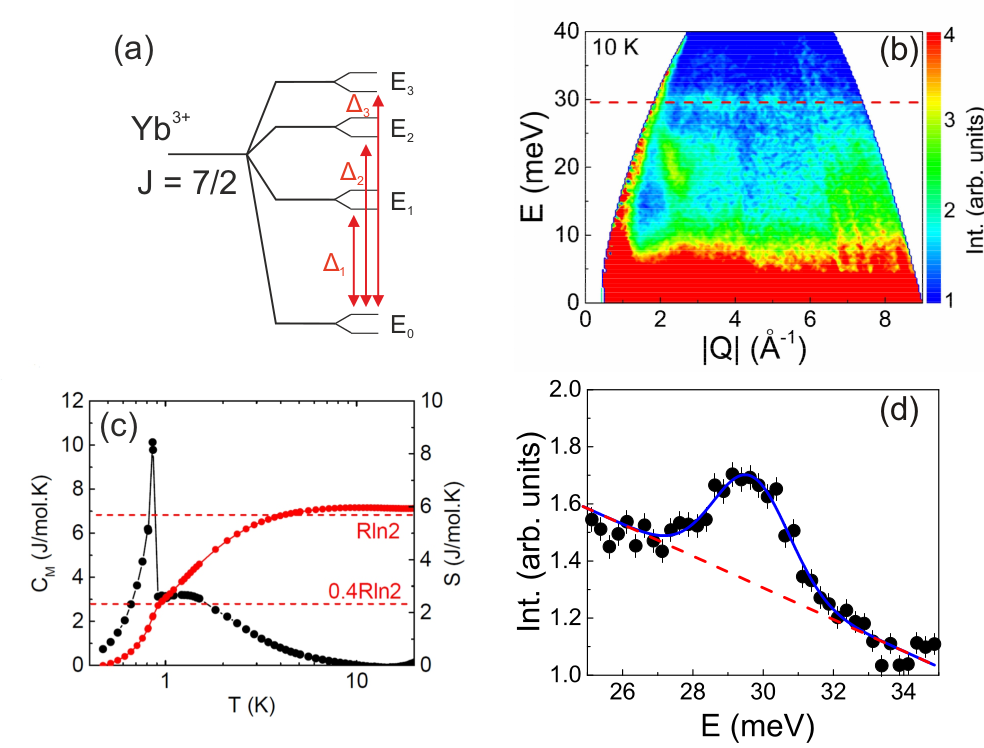}
  \caption{~Evidence for a CEF-induced pseudo-spin $S~=~1/2$ ground state. (a)~Sketch of the four isolated CEF doublet states of Yb$^{3+}$, where the eight-fold degeneracy of $J = 7/2$
  $(2J + 1 = 8)$ is lifted to four doublet states $E_0$, $E_1$, $E_2$, $E_3$, due to the low point symmetry. (b) INS spectrum of \YbAl\ measured at 10~K. A flat CEF mode is observed, indicated by the red dashed line. (c)~Low-temperature specific heat and integrated entropy of \YbAl. (d)~Energy dependent intensity integrated over the wave vector range $|Q| = [3, 7]$~\AA$^{-1}$.
  Reproduced with permission from~\cite{Wu2019PRB}.
}
  \label{Fig_S_one_half}
\end{figure}

To investigate the single-ion properties, the INS measurements were performed on the CNCS instrument and the results are shown in figure~\ref{Fig_S_one_half}. The spectrum measured at 10~K shows a flat, wave-vector-independent mode (red dashed line in figure~\ref{Fig_S_one_half}(b)), indicating the excitation from the ground state to the first excited CEF level. The energy-dependent intensity integrated over the wave-vector range $|Q|=[3$ -- $7]$~{\AA}$^{-1}$ is plotted in figure~\ref{Fig_S_one_half}(d).
The peak is found at $29.7\pm{0.1}$~meV, which corresponds to $\sim{345}$~K. This energy scale indicates a well separated ground-state doublet, which dominates the low-temperature magnetic properties.
This conclusion is further confirmed by low-temperature specific heat data. Figure~\ref{Fig_S_one_half} shows the temperature dependence of the magnetic specific heat, taken at $B = 0$~T, and the integrated entropy. A sharp peak anomaly is observed at $T_{\mathrm{N}}=0.88$~K in zero field, indicating the phase transition into a long-range ordered antiferromagnetic state. The entropy reaches $R\cdot\ln(2)$, as expected for the ground-state doublet at $\sim 5$~K, whereas only $\sim{0.45}{\;}R\cdot\ln(2)$ is released at $T_{\mathrm{N}}$, indicating the presence of strong magnetic fluctuations in the disordered phase. Since the experimental temperature ($T<5$~K) and magnetic field ($B<10$~T) discussed in the presented review are much smaller than the energy scale of the first excited CEF level $\Delta_{1}=29.7$~meV observed in INS measurements, the isolated ground-state doublet can be described as an effective spin $S={1/2}$.

\begin{figure}[tb!]
  \includegraphics[width=1\linewidth]{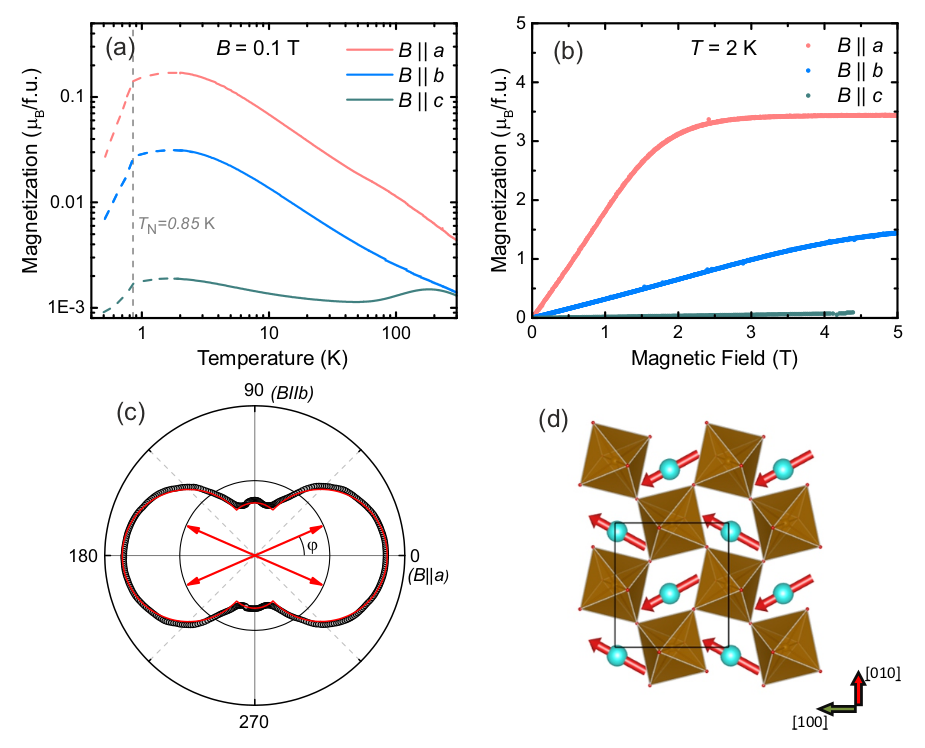}
  \caption{~Magnetic anisotropy of the ground state doublet in \YbAl.
  (a)~Temperature dependence of magnetization at $B = 0.1$~T.
  (b)~Field dependence of magnetization measured at $T = 2$~K.
  (c)~Magnetization as a function of angle collected at $T = 2$~K at magnetic field above the saturation, $B = 5$~T (see text).
  (d)~Crystal structure of \YbAl. The red arrows schematically show the easy-axis configuration of Yb$^{3+}$ with angle $\varphi = \pm{23.5}^{\circ}$ between the $a$ axis and the Yb magnetic moments.
  Reproduced with permission from~\cite{Wu2019Nat}.
}
  \label{Fig_anisotropy}
\end{figure}

Comprehensive magnetization measurements were performed in order to characterize the magnetic anisotropy of the ground-state doublet~\cite{Wu2019Nat}.
Figure~\ref{Fig_anisotropy}(b) shows the field dependence of the  magnetization measured along the three orthorhombic axes.
The high-field saturation moments along the $a$ and $b$ directions are more than one order of magnitude higher than the moment along the $c$-axis, confirming that CEF anisotropy constrains the Yb$^{3+}$ magnetic moments in the $ab$-plane.
A further measurement of the anisotropy in the $ab$-plane was performed with a horizontal rotator device.
The rotator allows one to continuously turn the sample along the axis, orthogonal to the field direction, and thus obtain the angular dependence of the magnetization at a given field and temperature.
As presented in figure~\ref{Fig_anisotropy}(c), the angular dependence of the magnetization, $M(\theta)$, measured in a magnetic field $B = 5$~T and $T = 2$~K has two minima at $\theta = 90\pm\varphi$, where $\varphi$ is the angle between the Ising moments and the $a$-axis, and the angle $\theta$ indicates the direction of the applied field in the $ab$-plane.
Assuming that the measured magnetization is a magnetization projection of the two Ising sublattices on the field axis, the angular dependence can be described as:

\begin{align}
	M&=\dfrac{M_{\mathrm{s}}}{2}(\lvert\cos(\varphi-\theta)\rvert+\lvert\cos(\varphi+\theta)\rvert) {\;},
	\label{Eq:MvsA}
\end{align}
where $M_{\mathrm{s}}$ is the saturation moment.
This analysis is shown as the red line in figure~\ref{Fig_anisotropy}(c).
With $M_{\mathrm{s}} = 3.8~\mu_{\mathrm{B}}\mathrm{/Yb}$, and $\varphi = 23.5^{\circ}$, the calculated curve matches perfectly well the experimental magnetization.
Rotating the coordinate system to align the quantization $z$-axis with the easy axis, the effective $g$-factors were estimated as  $g^{zz} = 7.6 \gg g^{xx} \simeq g^{yy}$ ($g^{xx} \approx 0.46$).
The value of the saturation moment as well as the canting angle $\varphi$ can be reasonably well reproduced by the PC model calculations (see section~\ref{cef-pcm} and \cite{Wu2019PRB}).
It is worth noting that the CEF anisotropy revealed by magnetization measurements and PC model calculations is responsible for the non-collinear magnetic structure of \YbAl, which was observed by means of powder neutron diffraction measurements~\cite{Radhakrishna1981}.

\subsubsection{Spin dynamics at zero field.}

\begin{figure}[tb!]
  \includegraphics[width=1\linewidth]{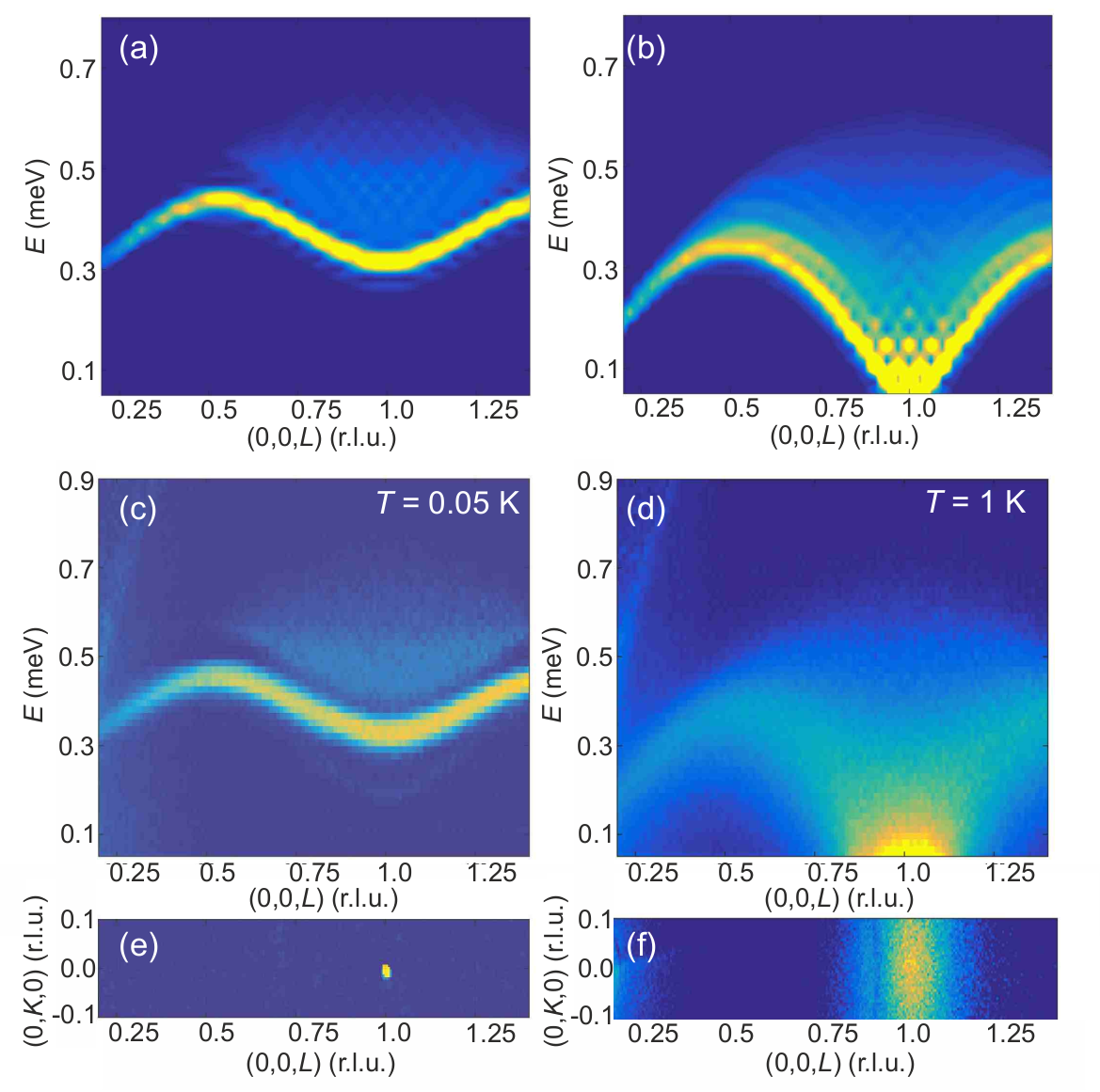}
  \caption{~DMRG-simulated (a,b) and observed (c,d) INS spectra along the $(00L)$ direction taken at $T = 0.05$ and 1~K ($E_{\mathrm{i}} = 1.55$~meV). The observed INS intensities were integrated within $K = [-2;2]$ and $H = [-0.3;0.3]$ in the orthogonal directions. (e,f) Contour plots of the magnetic scattering in the $(0KL)$ plane (integrated over the wave vector $H = [-0.1,0.1]$ r. l. u. and energy $E = [-0.1,0.1]$~meV), at $T$ = 0.05 and 1~K show a sharp (001) magnetic reflection and broad rod-shape diffuse scattering respectively. The intensity scales are different for each panel.
  Reproduced with permission from~\cite{Wu2019Nat}.
}
  \label{Fig_INS_ZeroField}
\end{figure}

The low-energy spin dynamics of \YbAl\ was studied by means of INS using the CNCS instrument.
Magnetic excitation spectra collected at $T = 0.05$ and 1~K, below and above $T_{\mathrm{N}}$, are shown in figure~\ref{Fig_INS_ZeroField}(c,d).
Gapless spinon excitations along the $(00L)$ direction are observed at 1.0~K, with a broad multi-spinon continuum, as it was predicted and observed in a number of $S = 1/2$ spin-chain magnets~\cite{lake2000, mourigal2013}, while the dispersion in the orthogonal directions is negligible.
Figure~\ref{Fig_INS_ZeroField}(f) presents a slice through the $(0KL)$ scattering plane at zero energy transfer. It shows a stripe-like feature elongated along the $K$-direction, indicating a significant correlation length along the chain direction in the paramagnetic phase.
No gap in the spin excitation spectrum can be resolved in the paramagnetic state at the magnetic Brillouin zone center $\mathbf{Q}=(001)$ within the instrumental resolution about of 0.05~meV.
Note that the temperature of the measurement, 1.0~K, is comparable to the instrumental resolution (0.05~meV), and that both are much smaller than the bandwidth of the two-spinon continuum, implying that we can safely exclude a scenario where the continuum arises from thermal or instrumental broadening.

With temperature decreasing below $T_{\mathrm{N}}$, the spectrum becomes gapped and splits into two modes: a low-lying resolution-limited intense magnon mode, and a weak continuum at higher energy. Moreover, the broad stripe-like feature at the elastic line is replaced by a sharp, resolution-limited Bragg peak, which is shown in figure~\ref{Fig_INS_ZeroField}(e).

Thus we conclude that at high temperature \YbAl\ is in a disordered, yet correlated state, due to a strong 1D intra-chain exchange interaction, while a finite interchain dipolar coupling causes conventional three-dimensional AFM order below 0.88~K.

\begin{figure*}[tb!]
  \includegraphics[width=0.95\linewidth]{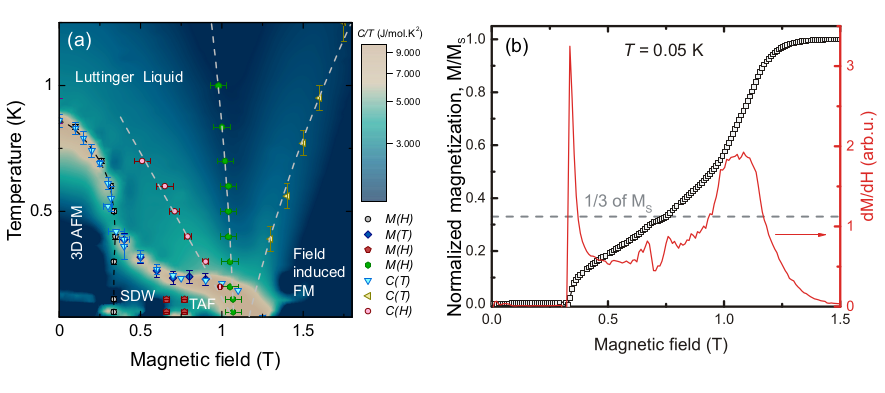}
  \caption{~(a) Magnetic field - temperature phase diagram of \YbAl. The contour plot shows the normalized magnetic specific heat, $C/T$. Magnetic phase boundaries extracted through different measurements are overplotted. (b)~Field dependence of the magnetization and its first derivative taken at $T = 50$~mK.
  Reproduced with permission from~\cite{Wu2019Nat}.
}
  \label{Fig_Phase_Diagram}
\end{figure*}

Note that the highly anisotropic ground state doublet of the Yb $J = 7/2$ multiplet was projected onto the $S = 1/2$ model, and the CEF-induced single-ion anisotropy is absorbed into an effective $g$-factor, which is included into the expression for the neutron cross-section~\eqref{exp_magnetic_cross_section}.
As was shown previously, $g_{zz} \gg g_{xx} \simeq g_{yy}$, and therefore, the longitudinal component $S^{zz}(\mathbf{Q},\hbar\omega)$ dominates the observed spectra, whereas the transverse components $S^{xx}$ and $S^{yy}$ are mainly hidden to the neutron scattering probe. This conclusion is supported by a polarization factor analysis of the INS spectra and by the spectral shape of the field-induced excitations, see ref~\cite{Wu2019Nat} and section~\ref{sec:YbAlO3_in_field}.

\subsubsection{Spin dynamics and magnetic phase diagram in field-induced states.}
\label{sec:YbAlO3_in_field}
The ground state and the magnetic excitations of a spin $S = 1/2$ Heisenberg magnet can be described using the Tomonaga-Luttinger liquid model, which exhibits gapless spinon excitations. The application of a magnetic field induces a quantum phase transition towards a field-polarized ferromagnetic phase and opens a gap in the spectrum~\cite{mikeska2004one, mourigal2013}. The temperature-field phase diagram for $B\parallel[100]$ was studied by means of magnetization, specific heat and neutron scattering~\cite{Wu2019Nat} measurements.

Let us consider the field dependence of the magnetization measured at $T = 50$~mK, which is shown in figure~\ref{Fig_Phase_Diagram}(b). At low field the magnetization is almost absent, but when the field increases above $B_{\mathrm{c1}} = 0.32$~T, the magnetization exhibits a sudden jump and starts to rapidly increase. There is a noticeable plateau at $M_{\mathrm{s}}/3$ at $B_{\mathrm{c2}} \approx 0.75$~T as clearly seen in the $dM/dB$ curve.
A further increase of the field changes the curvature of $M(B)$ at $B\approx{1.1}$~T, and then the magnetization saturates.
With temperature increasing, the first critical field $B_{\mathrm{c1}}$ remains unchanged up to 0.6~K and continuously decreases at higher temperatures, whereas the magnetization plateau at $B_{\mathrm{c2}}$ becomes unresolvable already at $T = 0.3$~K.

\begin{figure}[b]
  \includegraphics[width=1\linewidth]{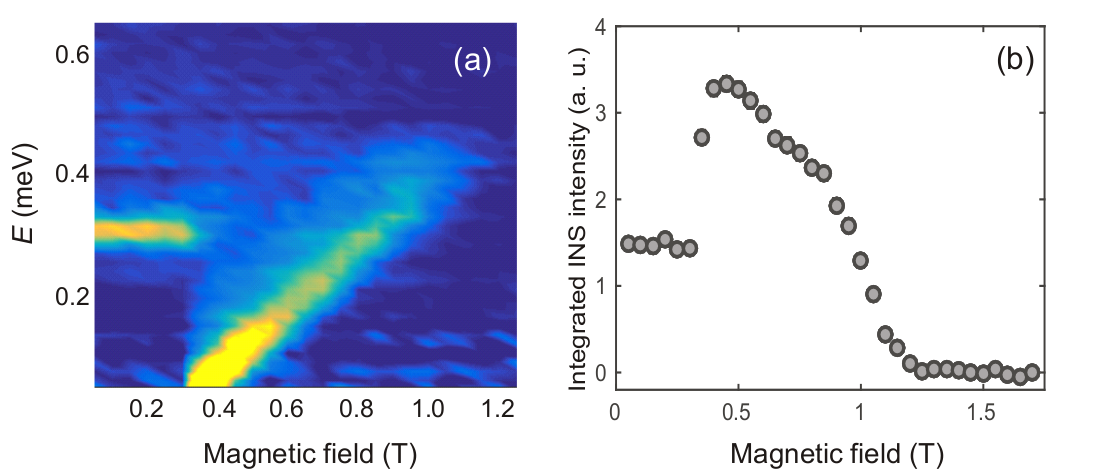}
  \caption{~(a)~INS intensity at $\mathrm{Q} = (001)$, $T = 0.05$~K plotted as function of energy transfer and magnetic field.
  (b)~Integrated INS intensity integrated over the whole $S(\mathrm{Q},\omega)$: $H = [-0.1,0.1]$, $K = [-0.1,0.1]$, $L = [0, 2]$ and $E = [0.05, 1]$~meV.
Reproduced with permission from~\cite{Wu2019Nat}.
}
  \label{Fig_INS_analysis_YbAlO3}
\end{figure}

The magnetic field - temperature phase diagram of \YbAl\ was reconstructed from the results of specific heat and magnetization measurements and is shown in figure~\ref{Fig_Phase_Diagram}.
The normalized specific heat $C/T$ was used as a contour plot, and the points represent phase boundaries and crossover temperatures extracted from the data.
One can distinguish several main regions: (i) the Luttinger-liquid phase at low field, dominated by strong 1D fluctuations above $T_{\mathrm{N}}$; (ii) the AFM commensurate phase at low temperature- and field-range below $T_{\mathrm{N}}$ and $B_{\mathrm{c1}}$ respectively; (iii) the field-induced spin density wave (SDW) phase with an incommensurate order parameter between the $B_{\mathrm{c1}} = 0.32$ and $B_{\mathrm{c2}} \approx 0.75$~T at temperatures below $\sim 0.3$~K; (iv) the transverse AFM phase (TAF) between $B_{\mathrm{c2}}$ and the quantum critical point (QCP); (v) the field-polarized FM phase at fields above the QCP; (vi) the region close to the QCP, dominated by quantum critical fluctuations.

\begin{figure*}[tb!]
  \includegraphics[width=0.9\linewidth]{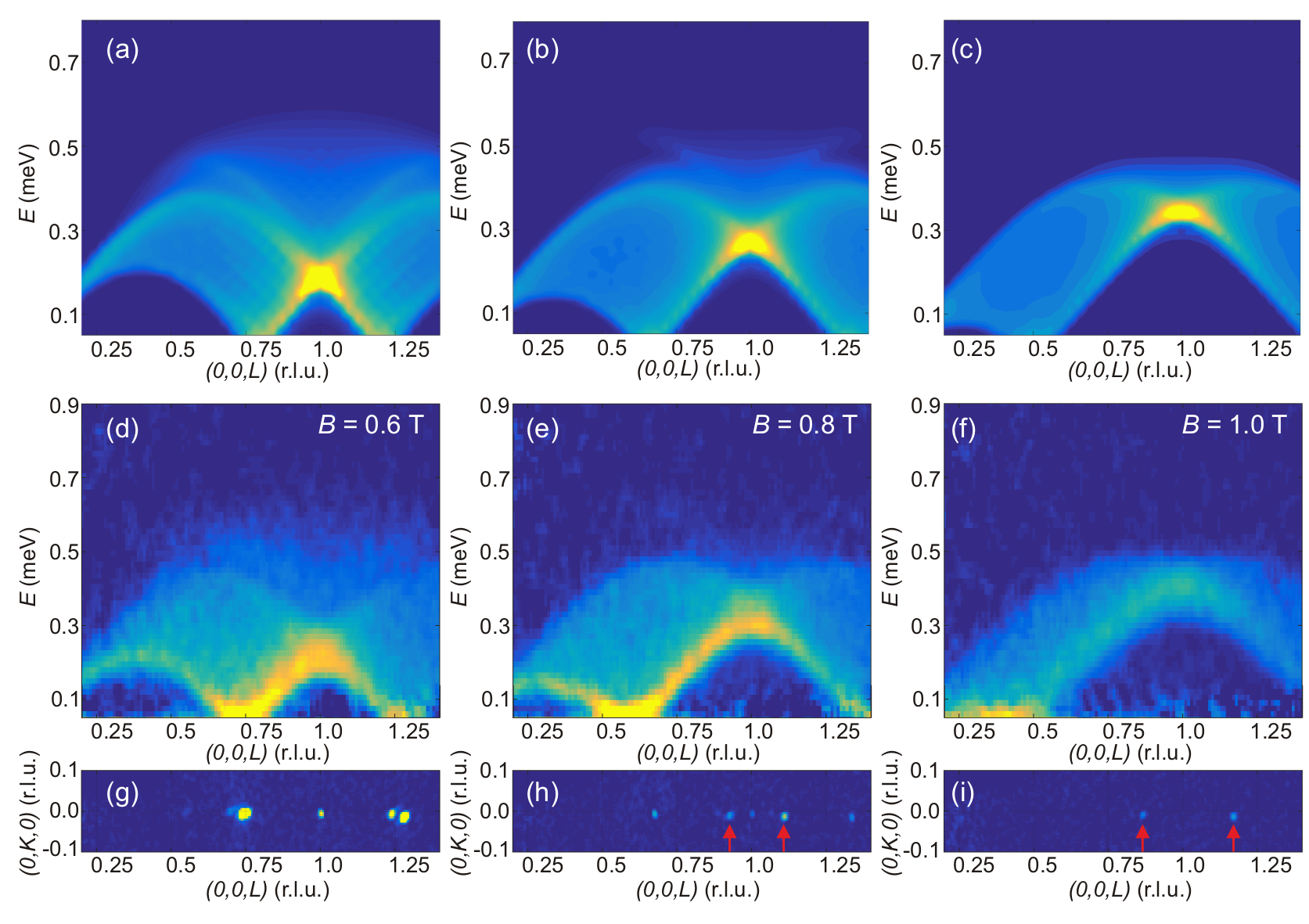}
  \caption{~DMRG-simulated (a,b,c) and observed (d,e,f) INS spectra along the $(00L)$ direction taken at $T = 0.05$~K and different magnetic fields, as indicated in the panels (($E_{\mathrm{i}} = 1.55$~meV). The observed INS intensities were integrated within $K = [-2;2]$ and $H = [-0.3;0.3]$ in the orthogonal directions. (g,h,i) Contour plots of the magnetic scattering in the $(0KL)$ plane integrated over wave vector $H = [-0.1,0.1]$ r. l. u. and energy $E = [-0.1, 0.1]$~meV at the same fields. Red arrows indicate incommensurate Bragg reflections, which originate from the twin. The intensity scale is different for each panel.
  Reproduced with permission from~\cite{Wu2019Nat}.
}
  \label{Fig_INS_inField}
\end{figure*}

The magnetic field - temperature phase diagram of \YbAl\ contains a number of different magnetic phases and thus naturally implies a nontrivial field-induced evolution of the magnetic excitations.
The field-induced excitations and magnetic structure of \YbAl\ were studied by means of neutron scattering using the CNCS instrument~\cite{Wu2019Nat}.

Figures~\ref{Fig_INS_analysis_YbAlO3} and~\ref{Fig_INS_inField} show the brief summary of the spectroscopic studies.
First of all, let us consider the field dependence of the INS signal at the $\Gamma$-point, $\mathbf{Q} = (001)$, and the INS intensity, integrated over the first Brillouin zone as a function of magnetic field, which are shown in figure~\ref{Fig_INS_analysis_YbAlO3}(a) and (b) respectively.
One can see that the application of the magnetic field up to $B_{\mathrm{c1}}$ does not affect the spectra. At $B_{\mathrm{c1}}$ the gap abruptly drops to zero. Further field increase causes a linear evolution of the spin excitations at $\mathbf{Q} = (001)$ up to the QCP. The overall spectral intensity behaves non-monotonically and gets fully suppressed at the QCP.

In the field region between $B_{\mathrm{c1}}$ and the QCP the spectrum becomes gapless (within the precision of the experimental resolution, $\sim 0.05$~meV), although the soft-mode position -- $\mathbf{Q}$-point, where the spectrum reaches zero -- continuously shifts away from $\mathbf{Q} = (001)$ towards incommensurate wavevectors, $\mathbf{Q} = (001\pm\delta_1)$, see figures~\ref{Fig_INS_inField}(d-f).
The incommensurability parameter of the soft mode, $\delta_1$, is directly proportional to the field-induced magnetization, $M/M_{\mathrm{S}}$, in agreement with the theoretical predictions for the Heisenberg spin-chain in a magnetic field~\cite{haldane1980general}.

In the field region between $B_{\mathrm{c1}}$ and $\approx 0.7$~T a sharp magnetic reflection was found directly below the minimum of the INS spectrum, see figure~\ref{Fig_INS_inField}.
In the field region between $\approx 0.85$~T and the QCP at 1.15~T the incommensurate reflections are absent although the sample remains in the magnetically ordered phase, as evident by sharp features in the thermodynamic measurements. This is suggested to be a transverse AFM state, see section~\ref{sec:YbAlO3_theory} for details.

\subsubsection{Theoretical description.}
\label{sec:YbAlO3_theory}

Having described the experimental finding we turn to the theoretical description of the unusual spin dynamics and phase diagram observed in \YbAl. According to the experimental results, a minimal model for \YbAl\ should consist of pseudo-$S={1/2}$ spin chains with isotropic or XY-like intra-chain interactions ($J \approx 2$~K), which are coupled via a long-range dipole-dipole interchain interaction. The estimate of the interchain dipolar coupling for the zero-field magnetic structure provides an energy scale of $\approx 0.67$~K per spin~\cite{Wu2019PRB}, consistent with the experimentally observed ordering temperature, $T_{\mathrm{N}} = 0.88$~K.
Moreover, the interchain dipolar interaction only couples the $z$-component of the spin operators, and the dipolar energy at the $i$-th spin can be rewritten as
$E_i^{\mathrm{inter}} \propto \sum_jS^z_iS^z_j$.
This makes \YbAl\ a rather unique example of a model material, in which the isotropic spin chains are coupled by an Ising-like interaction.

To model the magnetic behavior of \YbAl\ one has to take into account both interactions simultaneously, however it turns out to be a complex problem for numerical methods: On the one hand, calculations of long-range dipolar interchain couplings require the use of a large three-dimensional super-cell. On the other hand, numerical methods such as exact diagonalization or DMRG, which are able to simulate dynamics of quantum spin systems, deal with a large Hilbert space, which grows as $2^n$, where $n$ is the system size, and thus are ineffective for many-body three-dimensional systems.
Therefore, a ``clever'' assumption has to be made in order to simplify the problem.

So far, several theory approaches have been applied to describe the magnetic behavior of \YbAl.
In the simplest method the interchain coupling was replaced by an effective Zeeman field, $E_i^{\mathrm{inter}} \propto \langle{}S^z_j\rangle S^z_i = \vec{B}_i S_i^z$~\cite{Wu2019Nat} and in the commensurate AFM phase this term produces a so-called staggered field. This approximation reduces the problem to a single chain in a magnetic field, which was then solved by the DMRG method. The model zero-field spin Hamiltonian is given by:
\begin{align}
	\mathcal{H} &= J\sum_{i,j} ({S}^x_i{S}^x_j + {S}^y_i{S}^y_j + \Delta{S}^z_i{S}^z_j ) \nonumber \\
	&+ H_{\mathrm{st}}\sum_{i}(-1)^i{S}^z_i ,
 \label{Eq:StagField}
\end{align}
where $J$ is the isotropic exchange interaction and $\Delta$ parameterizes the exchange anisotropy
($\Delta = 0$ for the XY model, $\Delta = 1$ for the Heisenberg model, $J \rightarrow 0$ and $\Delta\cdot J = J_z$ for the Ising model).
First, let us consider the high temperature spectrum.
Above $T_{\mathrm{N}}$, $H_{\mathrm{st}} = 0$ because of the absence of a static ordered moment, and \eqref{Eq:StagField} becomes a simple XXZ model. Thus, the high-temperature spectrum can be used to refine the $J$ and $\Delta$ parameters.

It is well known that the XY and Heisenberg models exhibit gapless spectra, whereas for any $\Delta > 1$ there exists a finite gap.
As one can see in figure~\ref{Fig_INS_ZeroField}(d), within the experimental resolution of 0.05~meV one does not observe a gap above $T_{\mathrm{N}}$, which puts \YbAl\ to the XY or Heisenberg limit. Both models show similar dispersion, however the distribution of the spectral intensity changes considerably with $\Delta$. Fitting of the INS spectrum at 1~K along with the scaling of thermodynamic quantities in the vicinity of the QCP~\cite{Wu2019Nat} and measurements of the substituted sample, Y$_{0.96}$Yb$_{0.04}$AlO$_3$~\cite{Nikitin2020}, show that the anisotropy of the exchange interaction is surprisingly small, $\Delta \eqsim 1$, with $J = 0.21$~meV.

An elementary excitation of the Heisenberg model is a pair of spinons with $S = 1/2$, which propagate freely along the chain~\cite{mourigal2013}.
Below the ordering temperature, the dipolar interaction generates a finite staggered field, $H_{\mathrm{st}}$~\cite{alcaraz1995critical, zhou2021amplitude}. It produces a confining potential that increases linearly with the distance between both spinons, opens the gap in the spectrum, $\Delta_1 = (J*H^2_{\mathrm{st}})^{1/3}$ and splits the spectrum into a series of modes. Consequently, spinons get confined into bound states~\cite{Lake2010}. A fitting of the experimental spectrum has shown that $H = 0.055$~meV provides a good description of the observed data, see figures~\ref{Fig_INS_ZeroField}(c,d).
The spectra measured above $B_{\mathrm{c1}} = 0.35$~T exhibit gapless continua with a bandwidth of $\sim 0.5$~meV and are presented in figure~\ref{Fig_INS_inField} (d-f).
In order to describe the observed spectra, the XXZ Hamiltonian \eqref{Eq:StagField} with an additional Zeeman term was adapted.
Note that in the IC phase one cannot use the simple staggered field approximation because the $z$-component of the magnetization is incommensurate and therefore $M^z~\propto~\langle~S^z~\rangle~\mathrm{cos}(r\mathbf{q})$ and both, $\langle S^z \rangle$ and $\mathbf{q}$, depend on the magnetic field.
However, for $B_{\mathrm{c1}} < B < B_{\mathrm{QCP}}$, the Yb$^{3+}$ moments are strongly fluctuating even in the ordered phase.
As it was shown above (figure~\ref{Fig_S_one_half}), only about 45\% of the full $R\cdot\ln(2)$ entropy is released at the AFM transition at ambient field.
In the IC phase, this value decreases even further down to $\sim 20$\%, which indicates the strong decrease of the ordered moment~\cite{Wu2019Nat}.

Because the interchain molecular field is proportional to the magnitude of the ordered moment and the ordered moment within the IC phase is rather weak, one can ignore it in a first approximation.
This simplification reduces the Hamiltonian~\eqref{Eq:StagField} to the conventional Heisenberg model in an external field $H_{\mathrm{ex}}$.
Excitation spectra of this model were studied in detail previously~\cite{caux2005, caux2012}, and it was shown that the longitudinal and transverse spin structure factors look similar at $H = 0$ whereas at higher fields, their behaviors differ drastically.
Here, the DMRG calculations were used to obtain the longitudinal spin structure factor $S^{zz}(\mathbf{Q},\hbar\omega)$ above $B_{\mathrm{c1}}$, in various external fields $B_{\mathrm{ex}}/J = 0.8, 1.2, 1.6$, and the results are presented in figure~\ref{Fig_INS_inField}(a-c)\footnote{Note that the field values were used as fitting parameters for each spectrum separately, and they do not directly correspond to the external field, because an interchain interaction was not considered.}.
One can see a very good agreement between calculated and observed spectra, proving the longitudinal polarization of the observed excitations.
Furthermore, although a self-consistent treatment of the static moments from neighboring chains is required to obtain a more accurate excitation spectrum, an overall qualitative agreement with the experimental data is already achieved at this level of approximation.

\begin{figure}[tb!]
  \includegraphics[width=1\linewidth]{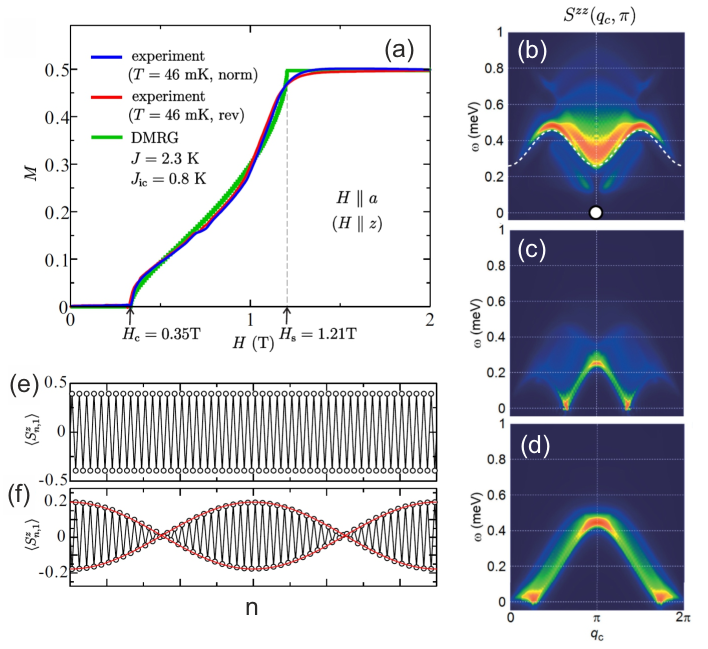}
  \caption{~DMRG modeling of a two-leg spin ladder. (a) Normalized magnetization of \YbAl\ measured at 46~mK and calculated using the Hamiltonian~\eqref{Eq:YbAlO3_DMRG}. (b-d) Longitudinal component of the dynamical spin structure factor, $S^{zz}(\mathbf{q},\omega)$ calculated at
  $B = 0$ (b) , 0.67 (c) and 1.03~T (d).
  (e-f) Expectation value of the $z$-projection of spin operator, $\langle{}S_n^z \rangle$ calculated at $B$ = 0~T (e) and within the SDW phase at $B = 0.36$~T (f).
  Reproduced with permission from~\cite{Agrapidis}.
}
  \label{DMRG}
\end{figure}

More advanced calculations were performed in \cite{Agrapidis}.
The authors considered a two-leg spin ladder, with a model Hamiltonian that is given by:
\begin{align}
	\mathcal{H} &= J\sum_i\sum_j {S}_{i,j}{S}_{i+1,j} + J_{\mathrm{ic}} \sum_i\sum_{j,j'}S^z_{i,j}S^z_{i,j'} \nonumber
	\\ &+H\sum_{i,j}{S}^z_{i,j},
 \label{Eq:YbAlO3_DMRG}
\end{align}
where the rung coupling, $J_{\mathrm{ic}}$ has only an Ising component, $J$ is the intra-chain leg coupling and $H$ is the  external magnetic field.
The model was studied with the DMRG technique. The calculated low-temperature magnetization curve was fitted to extract the model parameters, $J = 2.3$ and $J_{\mathrm{ic}} = 0.8$~K. Note that both parameters of inter- and intra-chain couplings are close to those deduced using the ``staggered field'' approach~\cite{Wu2019Nat}.
At zero field the ground state of the model is a simple commensurate AFM state. Above $H_{\mathrm{c1}} = 0.35$~T the model exhibits longitudinal SDW ordering in agreement with the experimental observation.
The expectation value for the local spin operator, $\langle S_i^z \rangle$, is plotted in figure~\ref{DMRG} (e-f) for AFM and incommensurate phases illustrating the transition between the two regimes. An increase of the magnetic field enhances the transverse correlation and quantum critical behavior is observed close to saturation.
Dynamical DMRG was applied to calculate the longitudinal component of the dynamical structure factor, $S^{zz}(\mathbf{q},\omega)$ and the resulting spectra are presented in figure~\ref{DMRG} (b-d). One can see that they capture the essential features of the observed spectra, i.e. it shows a finite gap at zero-field in the AFM state and gapless excitations above the $B_{\mathrm{c1}}$ critical field as well as a continuous field-induced evolution of the soft-mode.

Thus, despite the apparent oversimplification, the two-leg spin ladder model with Ising rung coupling surprisingly well reproduces the magnetism of \YbAl\ including the formation of an SDW phase, quantum critical behavior close to the QCP, unusual excitation spectra and the first-order phase transition between AFM and spin-density wave phases~\cite{Agrapidis}.

\begin{figure}[tb!]
  \includegraphics[width=1\linewidth]{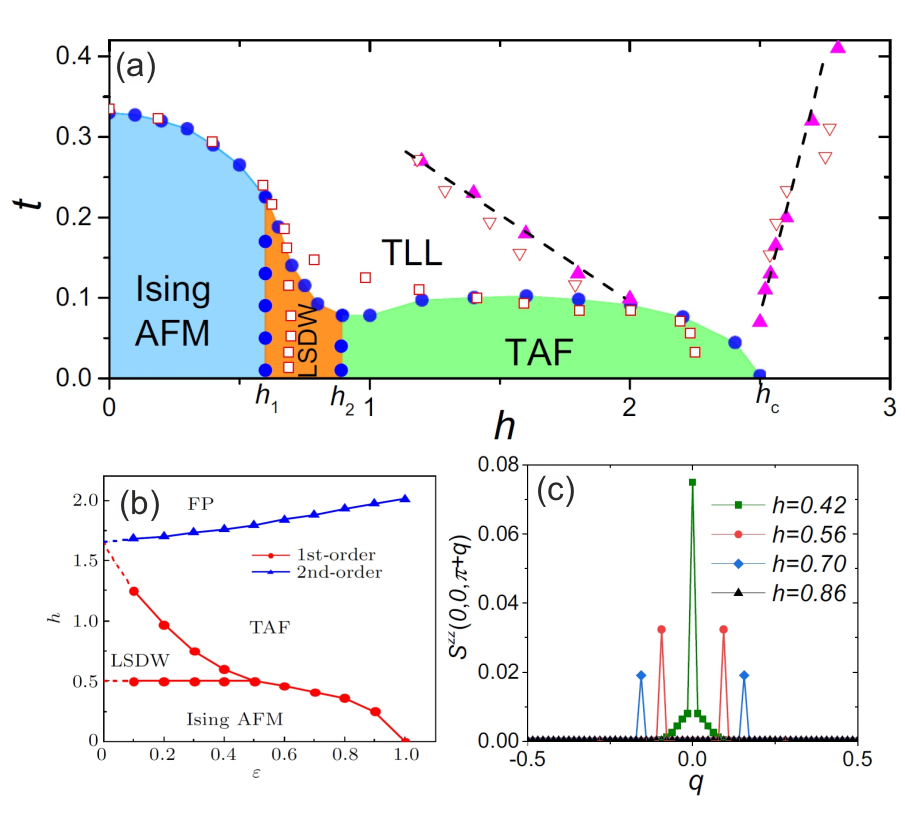}
  \caption{~Result of QMC modelling of the spin Hamiltonian~\eqref{Eq:YbAlO3_QMC}. (a) The magnetic field-temperature phase diagram of \YbAl\ with an Ising-like interchain exchange interaction, $\varepsilon\ = 0.25$. Filled (open) symbols denote calculated (experimentally determined) phase boundaries and crossover temperatures. (b)~The phase diagram of the Hamiltonian~\eqref{Eq:YbAlO3_QMC} as a function of magnetic field and exchange anisotropy. Panel (c) shows the longitudinal component of the static spin structure factor, $S^{zz}(0,0,\pi+q$), calculated within the SDW and TAF phases.
  Reproduced with permission from~\cite{Fan2020,Rong2020}.
}
  \label{Fig_QMC}
\end{figure}

An alternative approach was used in Refs.~\cite{Fan2020, Rong2020}.
The authors considered a three-dimensional spin lattice model with isotropic spin coupling along the chain direction and anisotropic XXZ interaction between the chains. The Hamiltonian is given by:
\begin{align}
	\mathcal{H} &= J_c\sum_i{S}_{i}{S}_{i+c}  - g\mu_{\mathrm{B}}H\sum_{i}{S}^z_{i} \nonumber \\
	&+ J_{ab} \sum_{i, \delta = \{a, b\}} [\epsilon(S_i^x S_{i + \delta}^x + S_i^y S_{i + \delta}^y) + S_i^z S_{i + \delta}^z ],
 \label{Eq:YbAlO3_QMC}
\end{align}
where $S_i = \{ S^x, S^y, S^z\}$ is a spin operator at site $i$, $J_c$ and $J_{ab}$ are intra- and inter-chain couplings and $\epsilon$ parameterized exchange anisotropy of the interchain coupling.
The authors studied the model using quantum Monte Carlo (QMC) simulations using a system with a maximal size of
$32 \times\ 32 \times\ 256$.

It is worth noting that Heisenberg spin chains coupled via a weak interchain interaction are a very well-known model system. At finite magnetic field, the transverse correlations always dominate the longitudinal ones, and consequently the canted TAF phase (sometimes also called a ``spin-flop phase''~\cite{yamashita1972field}) is expected to be the ground state at intermediate fields below the saturation. However, the QMC calculations show that the exchange anisotropy of the interchain coupling has drastic effect on the phase diagram~\cite{Fan2020}. Figure~\ref{Fig_QMC} (b) shows the phase diagram in $\epsilon$ - field coordinates. In the case of an  isotropic interchain coupling it consists of the AFM, TAF and FP (field-polarized) phases in agreement with simple expectations. However, one can see that in the case of a moderately strong Ising-like anisotropy, $\epsilon\ < 0.5$, the LSDW gets stabilized between the AFM and TAF regimes. A further reduction of $\epsilon$ expands the LSDW phase and is expected to completely suppress the TAF phase at the intermediate field regime in the $\epsilon = 0$ limit.

The experimentally determined phase diagram of \YbAl\ contains both LSDW and TAF phases, which means that $0 < \epsilon\ < 1/2$. The authors of \cite{Rong2020} fitted parameters of Hamiltonian~\eqref{Eq:YbAlO3_QMC} in order to reproduce the phase diagram of \YbAl\ and the calculated phase diagram and crossover points are shown in figure~\ref{Fig_QMC} (a) along with the experimental data. One can see that it consists of three long-range ordered phases. The first one is the low-field Ising AFM phase, which corresponds to a commensurate AFM state. The application of a magnetic field sequentially stabilizes longitudinal SDW and transverse antiferromagnetic (TAF) phases. The formation of the SDW phase manifests itself via sharp magnetic satellites at the spin structure factor, $S^{zz}(\mathbf{q}, \omega = 0)$, as shown in figure~\ref{Fig_QMC} (c).

We note that the formation of an SDW was also observed in other quasi one-dimensional spin systems, for example, BaCo$_2$V$_2$O$_8$ and SrCo$_2$V$_2$O$_8$~\cite{kimura2008longitudinal, shen2019magnetic}.
The appearance of an SDW phase in these materials is also related to the anisotropy of the exchange interactions, however the ultimate origin is related to the Ising-like anisotropy within the intra-chain exchange interaction, unlike the interchain coupling in \YbAl.

Summarizing, the magnetic phase diagram, the critical scaling of thermodynamic quantities, and the excitation spectra of \YbAl\ in a longitudinal magnetic field were studied by three different approaches and provide a fair description of the observed data. We note however, that each of those models has its own limitations owing to the approximations which were made to simplify the problem, and neither of them takes the long-range dipole-dipole interaction directly into account. However, all models consistently reproduce the essential features of the phase diagram of \YbAl\ including the formation of an SDW phase above $B_{\mathrm{c1}}$ as well as quantum critical behavior.

\subsection{Decoupled spin dynamics in \YbFe}
\label{YbFeO3}

The spin dynamics of the 3$d$ subsystem of \YbFe\ has been discussed in section~\ref{3d}.
The purpose of this section is to discuss experimental results concerning the spin dynamics of the Yb subsystem in the presence of ordered Fe moments.
It was shown that in $R$FeO$_3$ the influence of the Fe subsystem on the $R$ moment can be described in terms of an effective field~\cite{BelovBook, Belov1974, Belov1979magnetic}, using a modified mean-field theory approach~\cite{Bazaliy2004spin,Bazaliy2005, Tsymbal2007magnetic}.
Bazaliy \textit{et al.} analyzed a free energy functional of ErFeO$_3$~\cite{Bazaliy2004spin}.
They assumed that the ordered Fe subsystem polarizes nearly paramagnetic, strongly anisotropic moments of $R$-ions by an internal molecular field $\mathbf{H}^{\rm Fe}$.
In this model, one can take into account the influence of ordered Fe moments on the Yb subsystem with a simple Zeeman term and write down the general magnetic Hamiltonian for the Yb moments in the form
\begin{align}
 \mathcal{H}_{\rm Yb} &=  \sum_{l,m,i} B^l_mO^l_m(\mathbf{J}_i^{\rm Yb}) + J \sum_{i,j} \ {\mathbf{J}^{\rm Yb}_{i} \cdot \mathbf{J}^{\rm Yb}_{j}} \nonumber \\ &+ {\mathbf{H}}^{\rm Fe}\sum_i \mathbf{J}_i^{\rm Yb},
 \label{YbHam1}
\end{align}
where the first term is an one-site CEF Hamiltonian in Stevens notations~\cite{Stevens1952, Hutchings1964} (see section~\ref{cef-pcm}), the second term is the Yb-Yb inter-site Heisenberg exchange interaction, and the third term represents an influence of the Fe molecular field on the Yb magnetic subsystem.
Knowing that \YbAl\ and \YbFe\ have similar crystal structures, one can expect similar CEF and Yb-Yb exchange interactions in both materials.
Thus, the general Hamiltonian~\eqref{YbHam1}, which describes the magnetic behavior of Yb subsystem in \YbFe, presumably can be simplified to a $S=1/2$ spin-chain Hamiltonian in the magnetic field.

\subsubsection{Quantum quasi-1D excitations in the Yb subsystem of \YbFe\ in zero field.}

\begin{figure}[tb]
    \includegraphics[width=0.7\linewidth]{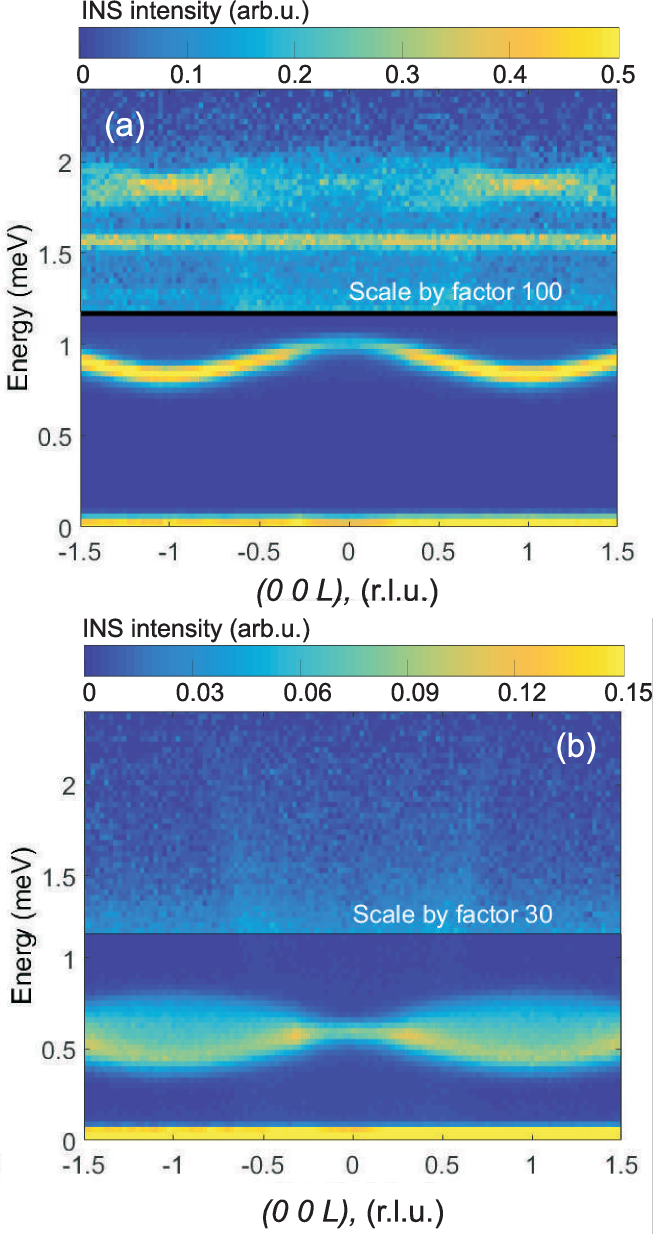}
    \caption{~Low-energy excitation spectra of \YbFe\ at $T=2$~K (a) and $T=10$~K (b) taken at CNCS. Energy slices are taken along the $(00L)$ direction with $(0K0)$ and $(H00)$ integrated over the range [-0.5, 0.5]. The intensity of the upper part of the panels has been scaled $\times 100$ and $\times 30$ to make two-kink excitations visible. The shadow in the center is an experimental artifact due to the direct beam. Reproduced with permission from~\cite{Nikitin2018}.}
    \label{YbFeO3_zerofield}
\end{figure}

The spin dynamics of the Yb subsystem in \YbFe\ was studied using the CNCS instrument with $E_{\rm i}=3$~meV.
Figure~\ref{YbFeO3_zerofield}(a,b) shows the experimentally observed intensity maps at temperatures below and above the SR transition.
The excitation spectrum  at $T=2$~K, below the SR transition, is dominated by a high-intensity sharp mode $E_1$, which disperses only along the $L$ direction.
We also observe a weak dispersionless excitation at $E_2\approx 1.5$~meV and a continuum centered at $E_3~\approx~1.8$~meV with dispersive boundaries and a bandwidth of $\Delta{}E\approx0.3$~meV at the zone center.
Above $T_{\mathrm{SR}}$, a bow-tie shaped continuum arises at $E\approx0.6$~meV with a sharp mode observed at the lower boundary, and the low-intensity excitation $E_2$ and the continuum $E_3$, present at $T=2$~K, disappear (or become much suppressed in intensity).
All observed excitations have negligible dispersion along other directions, indicating that the Yb moments form weakly coupled spin chains running along the $c$-axis despite the three-dimensional perovskite structure, similar to the isostructural \YbAl.
Note that in both spectra taken above and below $T_{\mathrm{SR}}$ a second ``shadow'' mode~\cite{Cabrera} with similar dispersion, but shifted periodicity, has been observed.
It has no intensity at $H = K = 0$, but becomes visible at larger $H$ and $K$.
This mode is associated with the buckling of the Yb chains~\cite{marezio1970crystal} (see the Supplementary Information of~\cite{Nikitin2018} for details).

In order to describe the low-energy magnetic excitations of the Yb$^{3+}$ moments observed in \YbFe, one can transform \eqref{YbHam1} into the one-dimensional XXZ $S=1/2$ Hamiltonian, similar to \eqref{Eq:StagField} used in section~\ref{sec:YbAlO3_theory}:
\begin{align}
 \mathcal{H_{\rm Yb}} &=  J_z \sum_i \ {S^z_i S^z_{i+1}} + J_{xy} \sum_i \ {(S^x_i S^x_{i+1} + S^y_i S_{i+1}^y)}\ \nonumber\\
 &+ \sum_i \mathbf{H_{\rm ef}} \cdot \mathbf{S}_i,
 \label{YbHam2}
\end{align}
where the first two terms correspond to the anisotropic exchange interaction between the nearest-neighbor Yb moments along the $c$ axis, and the last term is an effective Zeeman term -- the sum of the external field and the molecular field of the Fe subsystem.

At temperatures $T<T_{\mathrm{SR}}$ the net moment of the Fe subsystem is directed along the $a$ axis, as shown in figure~\ref{YBFO_Structure}(a), creating a \emph{longitudinal} field for the Yb$^{3+}$ spins.
The calculation of the eigenstates of \eqref{YbHam2} was done~\cite{Nikitin2018} using the zero-temperature exact diagonalization of a finite chain ($L=20$) with the ALPS software~\cite{ALPS1, ALPS2}.
A cosine-shaped dispersion of the lowest excitation with a maximum at the zone center suggests that the exchange interaction is antiferromagnetic and that the effective field $\mathbf{H}_{\rm ef}$ is large in comparison to $J_z$ and $J_{xy}$.
In this case all spins are parallel, $\langle S_n^z \rangle = S$~\cite{mourigal2013}.
The excitation spectrum of such a fully polarized state is similar to that of an FM chain and was discussed in various papers~\cite{Torrance595, Torrance587, Fogedby, Schneider, Orbach, Schneider1982}.
A single sharp mode with energy $E\approx1$~meV occurs due to scattering by a single-flip quasiparticle.
Besides, modes of an anisotropic FM or field-polarized AFM chain contain a two-kink bound state and a continuum consisting of pairs of independently propagating kinks.
The \YbFe\ data fit well to $J_z=-0.25$~meV, $J_{xy}=0.1$~meV and $\mathbf{H_{\rm ef}}=0.67$~meV.
The observed cross-section of two-kink states is about two orders of magnitude weaker than that for the single-flip excitation, in agreement with naive expectations~\cite{Torrance595}.

At temperatures $T>T_{\mathrm{SR}}$  the Fe net moment reorients along the $c$-axis, inducing a \emph{transverse} field for the Yb spins (see figure~\ref{YBFO_Structure}(c)).
However, at $T=10$~K the observed superposition of a bow-tie shaped continuum with a sharp excitation at the bottom (figure~\ref{YbFeO3_zerofield}(b)), suggests that the Yb sublattice is in a partially polarized state, as if a weak longitudinal field was still present.
A weak coupling between the magnetic chains in the $ab$ plane could be a possible explanation of the observed spectrum.
In a first approximation, such coupling can be replaced by an effective longitudinal mean-field \cite{Carr,Coldea}.
The spin-excitation spectrum in a skew ($H_x$, $H_z$) field is indeed characterized by a combination of a continuum due to scattering by pairs of kinks, which interpolate between regions with magnetization `up' and `down', and a sharp mode created by single spin-flip quasiparticles.
A finite temperature model of an XXZ chain is required to describe the details of the experimental spectra in this case.

\begin{figure*}[t!]
\includegraphics[width=0.8\linewidth]{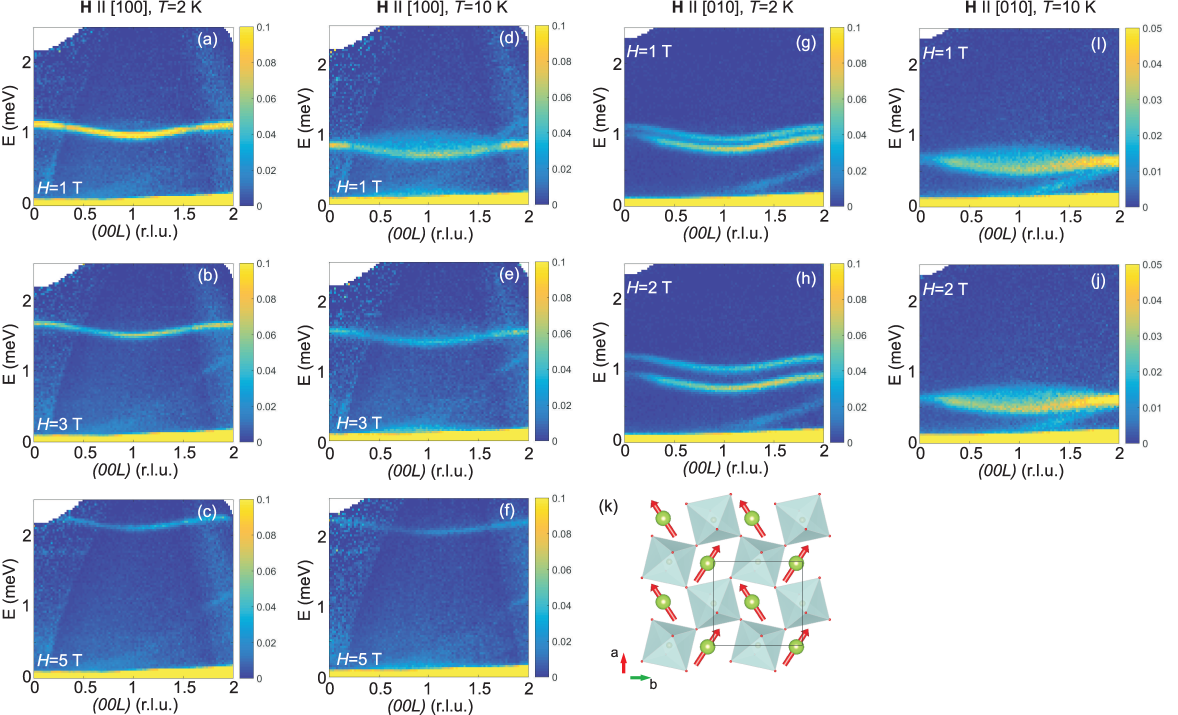}
  \caption{
  ~Effect of a magnetic field on the low-energy spin dynamics of \YbFe. The experimental spectrum along the $(00L)$ direction, measured at magnetic fields along the $a$ axis (a-c) and $b$ axis (g-j), at temperatures 2~K (a-c,g,h) and 10~K (d-f,i,j).  A visible linear diagonal line in the spectra is due to an instrumental effect.
  (k) Sketch of the field-induced magnetic structure of the Yb moments below $T_{\mathrm{SR}}$.
  Reproduced with permission from~\cite{Nikitin2018}.
  }
  \label{CNCS_magnetic}
\end{figure*}

\subsubsection{Effect of magnetic field on the low-energy spin dynamics of {\YbFe}.}

In this section we present the results of INS measurements with a magnetic field applied along all $a$, $b$ and $c$ axes of orthorhombic \YbFe.

In the low-temperature region ($T<T_{\mathrm{SR}}$), Yb spins are already polarized along the easy $a$ axis even without an external magnetic field.
The external field along the $a$-axis simply leads to an increased Zeeman splitting of the ground state, figure~\ref{CNCS_magnetic}(a-c).
At $T>T_{\mathrm{SR}}$, \YbFe\ is in the $\rm \Gamma2$ phase (see figure~\ref{YBFO_Structure}), and the net moment is directed along the $c$ axis.
Application of the magnetic field $\mathbf{H}\parallel{}[100]$ at $T=10$~K polarizes the Yb subsystem and induces an SR transition of the Fe-moments $\rm \Gamma4 \rightarrow{} \Gamma2$ at $H\approx4.3$~T~\cite{Nikitin2018}.
The INS spectrum exhibits larger Zeeman splitting, whereas the continuum, dominating at zero field, is rapidly suppressed and becomes undetectable already at $H=3$~T (see figure~\ref{CNCS_magnetic}(d-f)).
Above the field-induced SR transition, the spectra at both temperatures, $T=2$ and 10~K, become identical.

In contrast to the relatively simple case of $\mathbf{H}\parallel{}[100]$, a magnetic field applied along the $b$ axis qualitatively changes the excitation spectra.
At temperatures below $T_{\mathrm{SR}}$ the single-particle mode splits into two parallel modes (see figure~\ref{CNCS_magnetic}(g,h)), whereas above $T_{\mathrm{SR}}$, a magnetic field up to 2~T has a minor effect on the spectra (see figure~\ref{CNCS_magnetic}(i,j)).
Below $T_{\mathrm{SR}}$ the Yb moments have an Ising-like anisotropy, lie in the $ab$ plane with $\varphi \approx \pm{21}^{\circ}$ to the $a$ axis and are fully polarized by the molecular field of the Fe subsystem.
Schematically, the molecular-field-induced magnetic structure of the Yb subsystem below $T_{\mathrm{SR}}$ is shown in figure~\ref{CNCS_magnetic}(k).
The application of a magnetic field along the $b$ axis lifts the degeneracy between neighboring magnetic chains, increasing the fluctuation energy of Yb moments with a positive projection on the $b$ axis, $\varphi=+21^{\circ}$, and decreasing the energy for the opposite direction, $\varphi=-21^{\circ}$.
A further increase in field suppresses the energy of the lower mode down to zero with a simultaneous polarization of the Yb moments along the $b$ axis.

In \YbFe, Yb moments are coupled in chains running along the $c$ axis.
Therefore, in order to measure a dispersion along the $(00L)$ direction with a magnetic field applied along the $c$ axis both should be oriented parallel within the scattering plane and the triple-axis FLEXX instrument with the horizontal cryomagnet HM-1 was used for this purpose.

\begin{figure}
\includegraphics[width=0.9\linewidth]{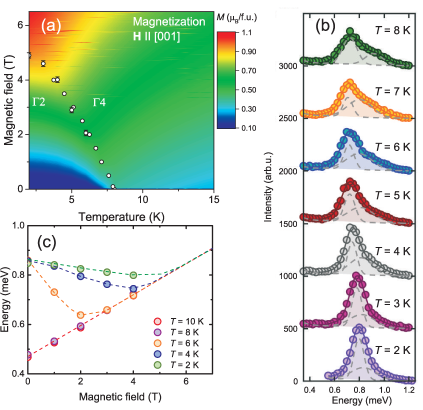}
  \caption{~Effect of magnetic field along the $c$ axis on the low energy spin dynamics of \YbFe.
  (a) Magnetic field - temperature phase diagram of \YbFe\ taken at $\mathbf{H}$ applied along the $c$-axis. The color-plot shows magnetization data.
  (b) Energy scans, measured with the FLEXX instrument at $H=4$~T and $\mathbf{Q}=$~(001) at various temperatures. The solid line is an overall fit of the magnetic signal. Dotted lines represent two Gaussian functions, used for the fitting.
  (c) Magnetic field dependence of the ``main mode'' peak as a function of magnetic field. Dotted lines are drawn to guide the eyes. Reproduced with permission from~\cite{Nikitin2018}.
}
  \label{Flexx}
\end{figure}

The low-$T$  $\rm \Gamma2$ phase is characterized by the weak net FM moment of Fe moments aligned along the $a$-axis. The magnetic field - temperature  phase diagram of \YbFe\ reconstructed from the magnetic measurements shows that application of magnetic field along the $c$-axis continuously suppresses the $\rm \Gamma2$ phase in favour of $\rm \Gamma 4$ (figure~\ref{Flexx}(a)) and the critical field $H_{\rm crit}^{\rm \Gamma2\rightarrow\Gamma4}$ gradually increases with decreasing temperature.
Inelastic spectra taken at $\mathbf{Q}=$~(001) and $H=4$~T are described by the combination of two modes, a resolution-limited intense peak (``main'' mode) and an additional broad peak at higher energy, see figure~\ref{Flexx}(b).
The low-temperature scans ($T=2$ and 3~K) in the $\rm \Gamma2$ phase show the largest contribution of the ``main'' mode.
The center of the second peak is located very close to the first one.
At $T=4$~K, a field-induced SR transition occurs.
The second peak shifts to higher energies and its intensity grows, whereas a further increase in temperature has no major effect on the spectra.

Figure~\ref{Flexx}(c) shows the magnetic field dependence of the ``main'' mode taken at different temperatures.
The excitation taken at $T=2$ and 4~K exhibits different behavior compared to that at  $T=6,8$ and 10~K.
In the low-temperature $\rm \Gamma2$ phase, an increasing field reduces the energy of the excitation until the critical field $H_c^{\rm \Gamma2\rightarrow\Gamma4}$ is reached (see figure~\ref{Flexx}(a)).
The excitation energy increases again at higher fields.
At temperatures above the SR transition, in the $\rm \Gamma4$ phase, the excitation energy always increases with field.

Summarizing this section, one can see that the effect of the external magnetic field on the spin dynamics is similar to that of the internal field by the Fe moments.
Because the low-energy excitations were found to have a dispersion along the $c$ axis only, we conclude that the Yb nearest neighbor AFM exchange interaction along the $c$ axis dominates the exchange interactions within the $ab$ plane, despite the 3D crystal structure of \YbFe.
A 1D XXZ $S=1/2$ Hamiltonian \eqref{YbHam2} with an additional Zeeman term describing the effective interaction with the Fe subsystem describes the Yb dynamics.
We observed an unusual transition between two regimes of the quasi-1D Yb fluctuations, induced by the rotation of the Fe molecular field, which serves as an intrinsic ``tuning parameter''.
We further note that although both Fe and Yb subsystems influence each other, their dynamics take place on essentially different energy scales. In addition, we did not observe collective magnon modes, which involve simultaneous excitation of both subsystems. Therefore, the spin dynamics of Fe and Yb subsystems is decoupled and was described considering effective spin Hamiltonians for Yb and Fe subsystems separately.

However, the currently available theoretical description of the excitation spectra in \YbFe\ cannot adequately capture all essential features of the measurements. In particular, the nature of the  continuum above the main mode, which was observed above $T_{\mathrm{SR}}$, remains unclear. Moreover, an exact diagonalization calculation shows that the two-magnon excitations observed in the experiment carry zero spectral weight in a simple XXZ model in magnetic field. Additional terms in the spin Hamiltonian~\eqref{YbHam2} are required to redistribute the spectral intensity towards these modes.
Moreover, fitting of the relative positions of the single magnon branch, two-magnon continuum and two-magnon bound state yields different signs for different components of the exchange interaction ($J_z$ is FM $J_{xy}$ is AFM), which is not a trivial result requiring theoretical justification.

\subsection{Low-energy excitations in \TmFe.}
\label{Chp:TmFeO3}

As we discussed in the Introduction, the details of the ground state of the 4$f$ magnetic sublattice strongly depend on whether the  rare-earth ion has an odd (Kramers ion) or even number of electrons in the partially filled 4$f$-shell.
In sections~\ref{YbAlO3} and \ref{YbFeO3} we have demonstrated that in the case of Yb$^{3+}$, a Kramers ion, the quantum states of the ground doublet, separated by a large energy gap from the first excited state, can be viewed as an effective spin-1/2.
As a result, the Yb magnetic sublattice exhibits enhanced quantum fluctuations.
In contrast, in the case of perovskite with the non-Kramers rare-earth ion, a singlet ground state is often stabilized due to the low symmetry of the CEF.
The low-energy spin fluctuations, and the coupling between the two spin subsystems (3$d$ and 4$f$) give rise to a variety of spectacular effects, like an unusual incommensurate phase in  \TbFe~\cite{Artyukhin} or inertia-driven spin switching in \HoFe~\cite{Kimel2009}.
Here, we discuss in more detail the low-energy spin dynamics of a non-Kramers rare-earth ion, using the example of \TmFe.

\TmFe\ crystallizes in an orthorhombic distorted perovskite structure (space group $Pbnm$) \cite{Tsymbal2006}.
The Fe magnetic sublattice orders antiferromagnetically at $T_{\mathrm{N}} \approx 632$~K \cite{Leake1968,Plakhty1983}, while the Tm sublattice does not order down to $T=1.6$~K~\cite{Leake1968, Skorobogatov2020low}.
Similar to other orthoferrites, the DM interaction induces a canting of the Fe magnetic moments and the spin reorientation transition $\Gamma4 \rightarrow \Gamma2$ takes a place with decreasing temperature at $T \approx 87\pm5$~K .

Now we turn to the description of the rare-earth subsystem. To address the CEF Hamiltonian~\eqref{Eq:CEF1} of the Tm ions in \TmFe, the PC model calculations were performed in~\cite{Skorobogatov2020low}, and the obtained set of $B_l^m$ parameters provides a reasonable description of the single-ion state. The low-symmetry CEF fully lifts the degeneracy of the ground state multiplet into 13 singlets and the first two excited levels are located at $E_1 = 1.9$~meV and $E_2 = 7.7$~meV. Only the $E_0 \rightarrow E_1$ and $E_0 \rightarrow E_2$ transitions have significant INS intensity at low temperatures ($k_{\mathrm{B}}T < E_1$), whereas  transition probabilities to another excited states are considerably weaker (for details, see Table~II in~\cite{Skorobogatov2020low}). The calculated energy of the first excited level is in agreement with the experimental value, $\sim 2.2$~meV, obtained by means of INS~\cite{Skorobogatov2020low}, optical spectroscopy \cite{malozemoff1970,malozemoff1971} and Terahertz time-domain spectroscopy~\cite{Zhang2016}, while the calculated position of the second excited level, 7.7~meV, deviates substantially from the observed experimental value of 4.8~meV. The calculated magnetization of the Tm moments is strongly anisotropic, with the easy axis pointing along the $[001]$ direction, and a saturation moment of $m_c \approx 5.5 - 6~\mu_{\mathrm{B}}$, in good agreement with the experimental observations.

\begin{figure}[tb!]
\includegraphics[width=1.0\linewidth]{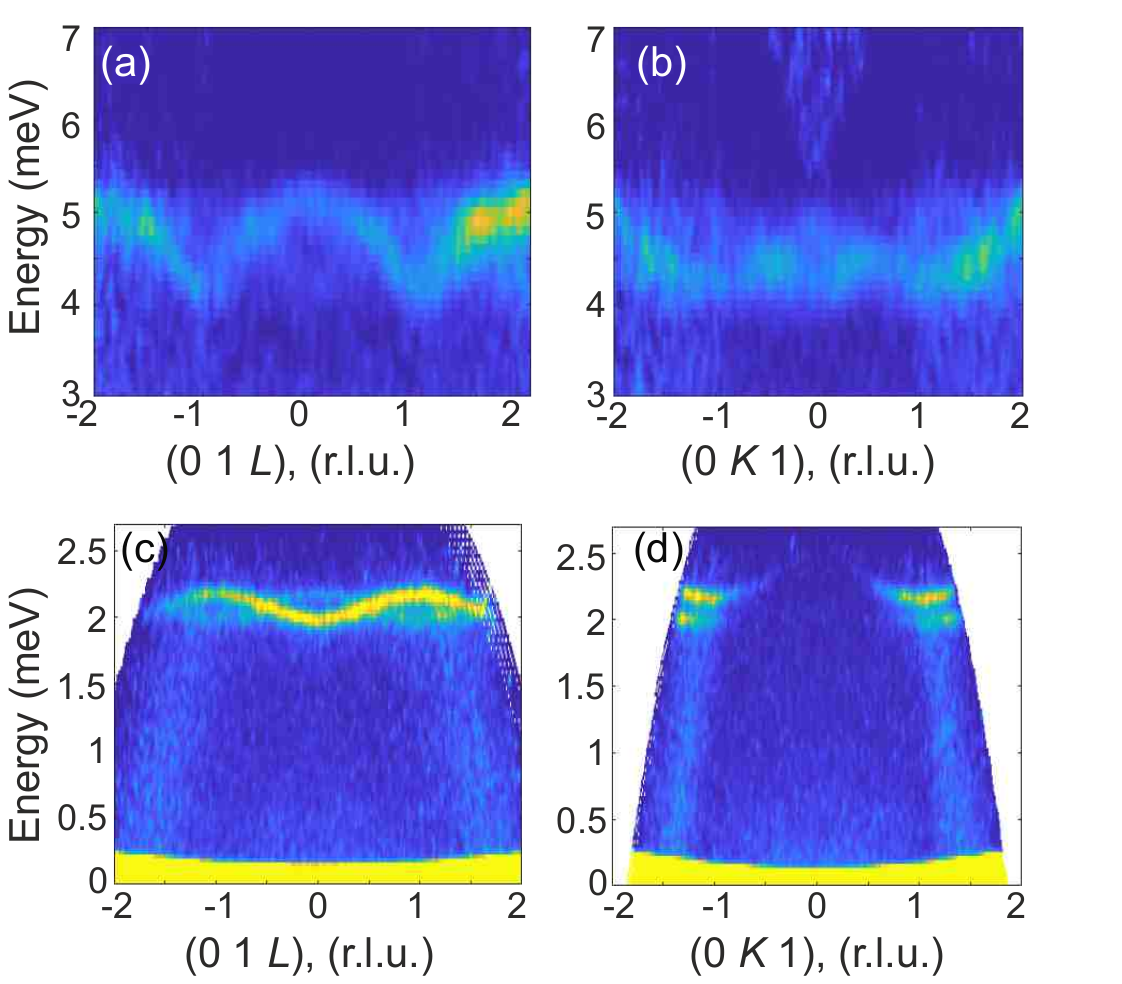}
\caption{~Low-energy excitation spectra of \TmFe\ measured in the $(0 1 L)$ (a,c) direction and the $(0 K 1)$ (b,d) direction at $T=1.7$~K. The spectra were taken with $E_{\mathrm{i}}=12$~meV (a,b) and $E_{\mathrm{i}}=3.3$~meV (c,d), and were integrated over 0.1~r.l.u. in the orthogonal directions. Reproduced with permission from~\cite{Skorobogatov2020low}.
}
\label{TmMagnon}
\end{figure}

Figure~\ref{TmMagnon} presents the important features of the low-energy INS excitation spectra of \TmFe.
Only two excitations located  at ${2.2}$ and ${4.8}$~meV have been observed at low-temperature regime ($k_{\mathrm{B}}T < E_1$ and energy transfer range $0 < \hbar \omega < 100$~meV. The observed excitations of Tm subsystem do not have quasi one-dimensional character, in contrast to the isostructural \YbAl\ (section~\ref{YbAlO3}) and \YbFe\ (section~\ref{YbFeO3}) materials discussed above. The dispersion along the $(0 1 L)$ direction consists of two cosine-like bands and is similar to the \YbFe\ case for the temperatures $T<T_{\mathrm{SR}}$ ~\cite{Nikitin2018}.
The dispersion of the ${4.8}$~meV excited level is more pronounced along the $L$-direction compare to the ${2.2}$~meV level and the dispersion bandwidth is larger, probably due to the larger effective moment of this singlet state. We further note that the second mode (``shadow mode''~\cite{Cabrera}), which has nonzero intensity at $K \neq 0, H \neq 0$ appears due to crystallographic buckling of the Tm sublattice.

\begin{figure}[ht!]
\includegraphics[width=0.9\linewidth]{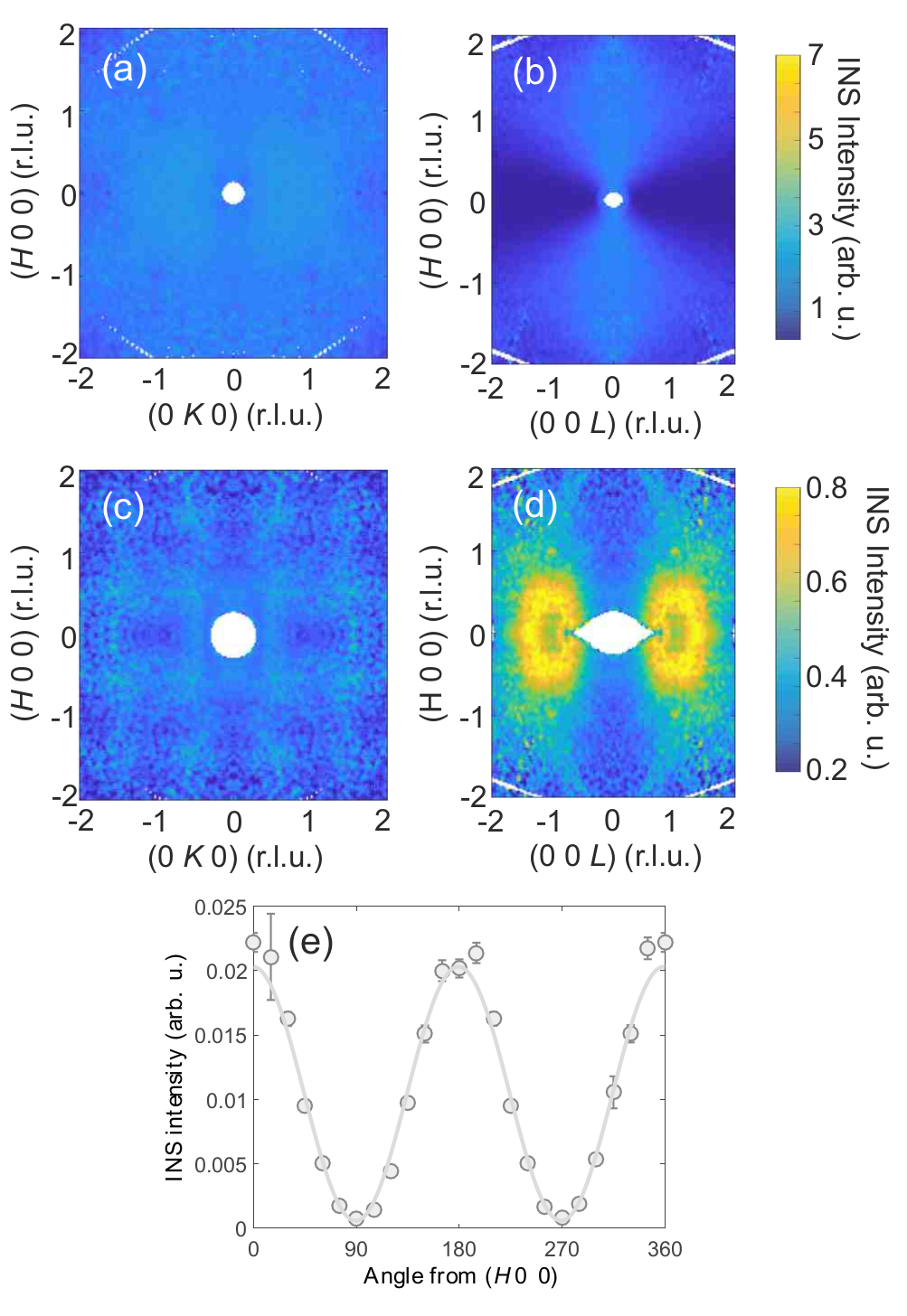}
\caption{~Constant energy slices of the INS intensity within the $(HK0)$ (a,c) and $(H0L)$ (b,d) planes taken at $T = 7$~K. The data were integrated within the energy windows of $E = [1.9,2.3]$~meV (a,b) and $E = [4.5,5]$~meV (c,d).
(e)~Angular dependence of the INS intensity in the $(H0L)$ scattering plane. The data are integrated within the energy window $E = [1.9,2.3]$~meV around a constant $|\vec{\xi}| = 1.13$~\AA$^{-1}$ with $\sim$ 0.05-0.07 (r.l.u) in the orthogonal directions. The solid line shows a fit with the $\cos^2(\theta)$ function. Reproduced with permission from~\cite{Skorobogatov2020low}.
}
\label{Polarization}
\end{figure}

Figure~\ref{Polarization} shows constant-energy slices within the energy range of two low-lying Tm excitations in the $(HK0)$ and $(H0L)$ planes. One can see that the excitation at ${2.2}$~meV has a strong asymmetry of the scattered intensity for the $(H0L)$ scattering plane, while remains isotropic in $(HK0)$. The observed behavior is similar to the polarization factor of neutron scattering, which was discussed in section~\ref{magscatt} and shown in  figure~\ref{Exp:ExpPolarization}. The polarization factor has a different form for transverse and longitudinal magnetic excitations:
\begin{align}
 P_{\mathrm{Long}} = (1-\vec{\xi}^2)   \label{eq:long}\\
 P_{\mathrm{Trans}} = (1+\vec{\xi}^2),  \label{eq:trans}
\end{align}
where $\vec{\xi}$ is a unit vector parallel to the directions of the magnetic moments. Note that the equations~\ref{eq:long} and \ref{eq:trans} produce qualitatively different INS patterns: the $P_{\mathrm{Long}}$ suppresses the INS intensity fully along a direction parallel to the magnetic moment, while $P_{\mathrm{Trans}}$ changes the intensity by not more than a factor of two. Figure~\ref{Polarization}(e) shows an angular dependence of the INS intensity integrated within the energy range of 2.2~meV excitation and $|\vec{\xi}|~\approx~1.13$~\AA$^{-1}$ in the $(0KL)$ plane. It can be fitted perfectly by a simple harmonic function, $I(\theta) = I_0\cdot \mathrm{cos}^2(\theta)$, where the $\theta$ is the angle between $\vec{\xi}$ and $(H 0 0)$ direction. The fitting shows that the INS intensity is suppressed almost completely for  $\vec{\xi} \parallel (0 0 L)$ in agreement with expectation for the longitudinal polarization, \eqref{eq:long}. Thus, the 2.2~meV excitation has a longitudinal polarization and is associated to the moment modulation along the $c$ axis. Note that this conclusion is further supported by the results of PC model calculations, which indicate considerable anisotropy of the matrix elements for the transition from the ground state to the first excited singlet with $\langle0|J^z|1\rangle \gg \langle0|J^x|1\rangle = \langle0|J^y|1\rangle = 0$. The PC model calculations also predict that the second transition, $E_0 \rightarrow E_2$, does not exhibit such an anisotropy in good agreement with the  isotropic INS patterns observed in experiment (figures~\ref{Polarization}(c-d)).

Usually, the description of the spin dynamics in localized magnets with a single-ion anisotropy starts from one of the limiting cases. If the magnetic anisotropy is weak, $J \gg\ K$, the LSWT can be applied. In the opposite regime, when the CEF splitting exceeds significantly the exchange interaction and the system has a Kramers doublet ground state, this doublet can be mapped onto the pseudo-$S = 1/2$ problem, whereas the CEF effect is taken into account by the anisotropic $g$-tensor, which, in general case, is unique for each doublet. Having the pseudo-$S = 1/2$ problem, one of the standard approaches  (LSWT, DMRG, exact diagonalization, etc.) can be applied to describe the low-energy spin dynamics of the doublet ground state~\cite{Wu2016Orbital, Wu2019Nat}.

The situation in \TmFe\ is more complex because the CEF splitting and the exchange interaction within the Tm subsystem are of comparable strength. Moreover, Tm and Fe subsystem interact with each other, which is evident by a presence of the SR transition~\cite{kimel2004laser, Tsymbal2006, Tsymbal2007magnetic} and considerable renormalization of the spin gap (up to $\approx 8$~meV~\cite{Skorobogatov2020low, Shapiro1974} at low temperatures),  which is $\approx 5-7$ larger compare to the isostructural compounds with non-magnetic $R$ ions~\cite{Park2017low, Hahn}.
Thus, it is crucial to disentangle Tm-Fe contribution from the interactions within the Tm subsystem, which can be done by measuring the magnetic excitations in an isostructural material with similar CEF, but nonmagnetic transition metal ion (e.g. TmAlO$_3$ or TmScO$_3$). Then, one can perform more sophisticated modeling of the spin dynamics taking into account both CEF and exchange, which are of the same order of magnitude. Therefore, further experimental and theoretical work is needed to fully resolve the microscopic spin Hamiltonian of \TmFe.

\section{Concluding remarks}
\label{outlook}

Inelastic neutron scattering has provided unique information on the spin dynamics of different magnetic subsystems in the orthoperovskite $RM$O$_3$ family of materials.
When both sublattices (3$d$ and 4$f$) are magnetic, the spectra consist of a superposition of two types of excitations, which coexist on much different energy scales.
Comparing the N\'{e}el temperatures of the rare-earth and transition-metal sublattices,  $T_{\mathrm{N}}^{3d} \sim 600$~K and $T_{\mathrm{N}}^{4f} \sim 1$~K, we expect a similar ratio of the exchange interactions $J_{3d-3d} \gg J_{4f-4f}$.
Indeed, the transition-metal spin dynamics are dominated by the high-energy gapped magnons with an energy scale of $\sim 60$~meV, while the excitation associated with the fluctuations of the rare-earth ion moments are limited to several meV or even less.
Since the transition-metal sublattice is magnetically ordered already at high temperatures, the LSWT is a good tool for the calculation of the $3d$ excitation spectra~\cite{Toth}.
In order to correctly describe the experimental data, the magnetic Hamiltonian includes the exchange interaction with antisymmetric DM terms and an anisotropy term \eqref{Eq:FeHam1}.
We note here that the details regarding the dispersion of magnetic modes, associated with $3d-4f$ exchange interactions, are mainly unexplored.

In this topical review we have mainly discussed unconventional low-energy magnetic excitations in $RM$O$_3$, which are impacted by the quantum effects of the anisotropic $R^{3+}$ moments.
The INS spectra are complicated and often consist of sharp magnon-like modes and a continuum, which is a fingerprint of multi-magnon or fractional spinon excitations.
The application of a magnetic field strongly perturbs the quantum ground state and can induce a quantum phase transition.

Over the past decade, there has been significant progress in the understanding of the fact that the spin-orbit entangled effective $S = 1/2$ moments of the rare-earth ions with ground state Kramers doublets generate enhanced quantum fluctuations and exhibit a variety of nearly degenerate states.
The ytterbium ion, Yb$^{3+}$ with an 4$f^{13}$ electron configuration, turns out to be especially favored~\cite{Li,Wu2016Orbital,Shen,Wu2019Nat,Bordelon}.
In particular, \YbAl\ discussed here, provides a realization of a quantum spin $S = 1/2$ chain material exhibiting both quantum critical Tomonaga-Luttinger liquid behavior and spinon confinement-deconfinement transitions.
An additional characteristic of \YbAl\ is that the moments can be saturated with a relatively low magnetic field of 1.1~T, enabling an exploration of the entire magnetic field-temperature phase diagram.

Clearly, more experimental and theoretical work is required to give a full description of the observed phenomena.
Currently, the exchange of the 4$f$ moments with the transition-metal sublattice was reduced to the effect of a molecular field.
An understanding the interplay between the $4f$ and $3d$ systems beyond the mean-field approximation may help to unveil the mechanism of the SR transition.
Additional insight is expected from measurements at high magnetic field, which may help to fully explore the magnetic phase diagrams.
For a correct description of the INS spectra at finite temperatures, and especially in the case of \YbFe\ at $T > T_{\mathrm{SR}}$, further theoretical work has to be done.

In addition, the crystal structure of $RM$O$_3$ naturally adapts to existing standard perovskite thin film substrates.
This can be exploited in future spinon-based research for material engineering under easily accessible laboratory conditions~\cite{Marchukov,Hirobe}.
Exploring a broader range of materials, like iron garnets or antiperovskites~\cite{Wang}, could also bring to light new unexpected physics.

We hope that the presented INS data and intriguing underlying physical phenomena will motivate further experimental and theoretical studies of the $RM$O$_3$ family of materials and renew the interest in the rich physics of rare-earth orthoferrites in general, along with other materials with a coexistence of several magnetic subsystems with different characteristic energy scales.

\section*{Acknowledgments}
We are grateful to Ch. R{\"{u}}egg for carefully reading the manuscript and for the useful comments.
We thank L. S. Wu, Z. Wang, W. Zhu, C. D. Batista, A. M. Tsvelik, A. M. Samarakoon, D. A. Tennant, M. Brando, L. Vasylechko, M. Frontzek, A. T. Savici, G. Sala, A. D. Christianson, M. D. Lumsden, S. A. Skorobogatov, K. A. Shaykhutdinov, A. D. Balaev, K. Yu. Terentjev, S. E. Hahn, G. E. Granroth, R. S. Fishman, A. I. Kolesnikov, E. Pomjakushina and K. Conder for their crucial contributions to joint studies related to the topics discussed here.
This research used resources at the Spallation Neutron Source, a DOE Office of Science User Facility operated by Oak Ridge National Laboratory.
S.E.N. acknowledges funding from the European Union’s Horizon 2020 research and innovation program under the Marie Sk{\l}odowska-Curie grant agreement No 701647.

\bibliographystyle{iopart-num-long-doi}
\bibliography{biblio}

\end{document}